\documentclass[pdflatex,sn-mathphys-num]{sn-jnl}

\usepackage{longtable}
\usepackage{graphicx}
\usepackage{multirow}
\usepackage{amsmath,amssymb,amsfonts}
\usepackage{amsthm}
\usepackage{mathrsfs}
\usepackage[title]{appendix}
\usepackage{xcolor}
\usepackage{textcomp}
\usepackage{manyfoot}
\usepackage{booktabs}
\usepackage{algorithm}
\usepackage{algorithmicx}
\usepackage{algpseudocode}
\usepackage{listings}
\usepackage{rotating}
\usepackage{pdflscape}
\usepackage{tabularx}
\usepackage{subcaption}
\usepackage{float}

\theoremstyle{thmstyleone}

\theoremstyle{thmstyletwo}

\theoremstyle{thmstylethree}
\newtheorem{definition}{Definition}

\begin{document}
	
	\title[Article Title]{A SAT-Based Exact Approach for Radio $k$-Labeling}
	
	\author[1,2]{\fnm{Huong} \sur{Vu Thanh}}\email{25028017@vnu.edu.vn}
	
	\author[1]{\fnm{Duc} \sur{Dao Van}}\email{24021409@vnu.edu.vn}
	
	\author*[1]{\fnm{Khanh} \sur{To Van}}\email{khanhtv@vnu.edu.vn}
	
	\affil[1]{\orgname{VNU University of Engineering and Technology}, \orgaddress{\city{Hanoi}, \country{Vietnam}}}
	
	\affil[2]{\orgname{Vietnam Maritime University}, \orgaddress{\city{Hai Phong}, \country{Vietnam}}}
	
	\abstract{The radio $k$-labeling (or $k$-coloring) problem seeks a minimum-span assignment of nonnegative integer labels to the vertices of a connected graph $G$ such that $ |f(u)-f(v)| \ge k+1-d(u,v) $ for all vertex pairs. Although numerous theoretical constructions and some heuristic algorithms have been proposed, existing approaches generally fail to provide certified optimal solutions for broad graph classes.
This paper presents an exact SAT-based framework for radio k-labeling that combines a compact order encoding with incremental SAT solving. The proposed framework incrementally tightens the admissible span while reusing learned clauses across SAT calls, avoiding repeated formula reconstruction. Experimental results on 146 benchmark instances from nine graph families demonstrate that the proposed approach establishes 38 new best-known radio numbers while matching or improving the best-known radio numbers on 130 of the 146 benchmark instances. 
The proposed SAT framework outperforms state-of-the-art commercial optimization solvers, including CPLEX and Gurobi, in terms of overall solution quality, and substantially improves upon previously published heuristic methods.
Furthermore, by combining the SAT frameworks with ILP models solved by CPLEX and Gurobi, the study certifies optimal solutions for 109 of the 146 benchmark instances, substantially expanding the set of radio-labeling benchmarks with proven optimality. These results demonstrate the effectiveness of incremental SAT solving as a practical exact optimization framework for difficult graph-labeling problems.
	}
	
	
	\maketitle
	
	\section{Introduction}
	\label{intro}
	
	The channel assignment problem, introduced by Hale~\cite{Hale1980}, models the allocation of frequencies to radio transmitters while minimizing interference and total bandwidth usage.
	In graph-theoretic formulations, vertices represent transmitters and edge or distance relations represent interference constraints between
	pairs of transmitters.
	This problem has motivated a wide range of graph labeling models, including the well-known $L(2,1)$-labeling introduced by Griggs and
	Yeh~\cite{GriggsYeh1992}, where adjacent vertices must receive labels at least two units apart and vertices at distance two must receive distinct labels.
	
	Computing optimal radio labelings is computationally challenging because every pair of vertices potentially introduces a distance-based constraint. Although many theoretical constructions and heuristic algorithms have been proposed for specific graph families, exact solutions remain known only for limited classes of highly structured graphs ~\cite{LiuZhu2005,Liu2008,saha2012graph,badr2020upper}.
	For more general graphs, most existing methods provide only feasible solutions or analytical bounds without certifying optimality~\cite{KolaPanigrahi2015,Das2017}.
	Integer linear programming (ILP) formulations can compute exact solutions in principle, but their scalability is often limited by the rapid growth of the search space~\cite{badr2020upper}.
	
	Boolean satisfiability (SAT) solving is one of the most successful paradigms for solving difficult combinatorial optimization problems~\cite{biere2009handbook}.
	Modern SAT solvers combine clause learning, conflict analysis, restarts, propagation techniques, and sophisticated branching heuristics, enabling them to solve instances containing millions of variables and clauses~\cite{marques1999grasp,moskewicz2001chaff}.
	SAT-based methods have achieved strong performance in scheduling, verification, planning, graph problems, and combinatorial design~\cite{biere2009handbook}.
	In particular, incremental SAT solving allows a sequence of closely related formulas to be solved while preserving learned clauses across solver calls, significantly reducing redundant computations~\cite{queue2019cadical}.
	These developments motivate the use of SAT solving for the radio $k$-labeling problem.
	The radio constraints are inherently combinatorial and involve a large number of pairwise distance conditions, making them suitable for Boolean encoding.
	Moreover, the optimization objective can be naturally handled through incremental tightening of the admissible span bound.
    Despite extensive research on theoretical constructions, analytical bounds, and heuristic algorithms, exact computational approaches remain scarce. To the best of our knowledge, no general SAT-based exact optimization framework has been proposed for the radio k-labeling problem.
	
	In this paper, we present an exact SAT-based framework for radio $k$-labeling that combines an order-encoding representation of vertex labels with an incremental SAT solving strategy. The SAT model is constructed once and progressively strengthened by iteratively forbidding previously obtained span values, reusing learned clauses across solver calls rather than rebuilding the formula from scratch. This design allows the framework to compute certified optimal spans within practical running times on many benchmark instances and, even when the optimality proof cannot be completed within the time limit, to still return feasible solutions that are consistently tighter than existing heuristic and ILP-based methods.	
		
		
		
    The main contributions of this work are summarized as follows:

\begin{itemize}

\item We propose an exact SAT framework for the radio $k$-labeling problem that combines a compact order encoding with an incremental optimization strategy. The framework progressively tightens the admissible labeling span while reusing learned clauses across successive SAT calls, eliminating the need to rebuild the SAT formula.

\item We establish 38 new best-known radio numbers among the 98 benchmark instances without a closed-form optimum, and on the 48 path and cycle instances, whose optimal radio number is already known in closed form, the SAT framework independently recovers the proven-optimal value on 46 instances.
 
\item By combining the proposed SAT frameworks with ILP models solved by CPLEX and Gurobi, we certify optimal solutions for 109 benchmark instances, substantially enlarging the collection of radio $k$-labeling benchmarks with proven optimality.

\item More broadly, this work demonstrates that modern incremental SAT solving constitutes a practical exact optimization paradigm for graph labeling problems that have traditionally been dominated by theoretical constructions and heuristic algorithms.

\end{itemize}
	
	The remainder of the paper is organized as follows.
	Section~\ref{sec:prel} introduces the necessary preliminaries and the ILP formulation.
	Section~\ref{subsec:relatedworks} reviews related works on radio labeling.
	Section~\ref{sec:SAT} presents the SAT encoding and the incremental solving framework.
	Experimental results are reported in Section~\ref{sec:exp&result}, followed by concluding remarks.

	\section{Preliminaries}
	\label{sec:prel}
	
	\subsection{Basic Definitions and Notation}
	\label{subsec:notations}
	
	Let $G=(V,E)$ be a finite, undirected, connected simple graph with $|V|=n$ vertices and $|E|=m$ edges. For any two vertices $u,v\in V$, let $d(u,v)$ denote the length of a shortest path between $u$ and $v$. The diameter of $G$ is defined as
	$
	\mathrm{diam}(G)=\max_{u,v\in V} d(u,v).
	$
	
	Throughout this paper, we consider only the standard radio-labeling setting
	$ k=\mathrm{diam}(G). $
	Consequently, the problem studied in this work corresponds to the classical radio-labeling problem.
	This choice is motivated by two complementary reasons.
	First, setting $k = \mathrm{diam}(G)$ yields the most constrained instance of the radio $k$-labeling problem: since $d(u,v) \le \mathrm{diam}(G)$ for every pair of distinct vertices, the required separation $\Delta_{uv} = k+1-d(u,v)$ is strictly positive for \emph{all} $\binom{n}{2}$ vertex pairs, so no pairwise radio constraint is trivially satisfied.
	For smaller values of $k$, pairs at distance greater than $k$ contribute no constraint, strictly reducing the constraint density and generally making the problem easier to solve.
	Second, this setting coincides with the classical radio labeling problem studied extensively in the literature~\cite{LiuZhu2005,Chartrand2001}, enabling direct comparison with existing exact results, heuristic upper bounds, and ILP formulations on standard benchmark families.
	The exact optimization formulations developed in this paper (Section~\ref{sec:SAT}) apply directly to arbitrary values of $k$ by substituting the desired parameter into the encoding of Section~\ref{sec:SAT}; the restriction to $k = \mathrm{diam}(G)$ is adopted here solely to focus the experimental evaluation on the hardest and most widely studied case.
	
	Since $ d(u,v)\le \mathrm{diam}(G)=k $
	for all distinct vertices $u,v\in V$, every pair of vertices induces a nontrivial radio constraint. For convenience, define the required label separation
	$ \Delta_{uv}=k+1-d(u,v). $
	Then $ \Delta_{uv}\ge1 $
	for every distinct pair of vertices.
	
	\begin{definition}[Radio labeling]
		A radio labeling of a graph $G$ is a mapping
		$
		f:V\rightarrow \{0,1,2,\ldots\}
		$
		such that
		$
		|f(u)-f(v)|\ge k+1-d(u,v)
		$
		for every pair of distinct vertices $u,v\in V$, where
		$
		k=\mathrm{diam}(G).
		$
	\end{definition}
	
	The radio constraint requires vertices that are close in the graph to receive labels that are sufficiently far apart numerically. In particular, adjacent vertices require the largest label separation.
	
	\begin{definition}[Span]
		The span of a radio labeling $f$ is defined as
		$
		\mathrm{span}(f)=\max_{v\in V} f(v).
		$
	\end{definition}
	
	\begin{definition}[Radio number]
		The radio number of $G$, denoted by $rn(G)$, is the minimum span among all feasible radio labelings of $G$:
		$
		rn(G)=\min_f \mathrm{span}(f).
		$
		Any labeling achieving this minimum value is called an optimal radio labeling.
	\end{definition}
	
	The radio-labeling problem can therefore be viewed as a combinatorial optimization problem whose objective is to minimize the maximum assigned label while satisfying all pairwise distance constraints.
	
	A classical characterization of the radio number is based on vertex orderings. Let
	$ (x_1,x_2,\ldots,x_n) $
	be an ordering of the vertices of $G$. If consecutive label differences satisfy
	$$ f(x_{i+1})-f(x_i)=k+1-d(x_i,x_{i+1}),$$
	then the resulting span can be expressed as
	$$ rn(G) = (n-1)(k+1)- \sum_{i=2}^{n} d(x_i,x_{i-1}) +1. $$
	
	This characterization shows that minimizing the radio span is closely related to maximizing the accumulated distance between consecutive vertices in a vertex ordering. Such formulations play an important role in many theoretical constructions developed in the literature. However, constructing optimal orderings becomes increasingly difficult for general graphs, motivating the development of exact optimization formulations such as ILP and SAT encodings.
	
	\subsection{ILP Model}
	\label{subsec:ilp}
	
	The study of Badr and Moussa~\cite{badr2020upper} formulated the radio $k$-coloring problem as an integer program with an integer variable $L_v$ denoting the label of vertex $v$, absolute-value pairwise constraints, and objective $\min \sum_v L_v$. This objective does not capture the radio number, minimizing the sum of labels is not equivalent to minimizing the maximum label, since a labeling with small sum may still have large maximum, and conversely; the correct objective is $\min\max_v L_v$.

    To obtain an exact baseline that targets the correct objective and shares its variables with the SAT encodings of Section~\ref{sec:SAT}, we use a binary formulation, referred to as \emph{ILP}; the baseline reported throughout is solved with CPLEX and denoted \emph{ILP CPLEX}.
	
	\begin{equation}
		x_{v,l} =
		\begin{cases}
			1, & \text{if vertex } v \text{ receives label } l, \\
			0, & \text{otherwise,}
		\end{cases}
		\qquad \forall\, v \in V,\ l \in \{1,\dots,\mathrm{UB}\},
		\label{eq:ilp-x}
	\end{equation}
	the single integer span variable
	$l_{\max} \in \mathbb{Z}_{\ge 0}$. The label of a vertex is recovered as
	$f(v) = \sum_{l=1}^{\mathrm{UB}} l\,x_{v,l}$.
	
	The ILP model (P) is
	\begin{align}
		\min \quad & l_{\max}
		\tag{P} \label{eq:ilp-obj} \\[3pt]
		\text{s.t.} \quad
		& \sum_{l=1}^{\mathrm{UB}} x_{v,l} = 1
		&& \forall\, v \in V,
		\label{eq:ilp-assign} \\[2pt]
		& l\,x_{v,l} \le l_{\max}
		&& \forall\, v \in V,\ l \in \{1, 2, \ldots, \mathrm{UB}\},
		\label{eq:ilp-lmax} \\[2pt]
		& x_{u,a} + x_{v,b} \le 1
		&& \forall\, u \ne v,\ \forall\, (a,b) : |a-b| < \Delta_{uv},
		\label{eq:ilp-radio} \\[2pt]
		& x_{v,l} \in \{0,1\},\quad
		l_{\max} \in \mathbb{Z}_{\ge 0}.
		&&
		\label{eq:ilp-dom}
	\end{align}
	
	Constraint~\eqref{eq:ilp-assign} assigns exactly one label to each vertex.
	Constraint~\eqref{eq:ilp-lmax} forces $l_{\max}$ to be at least the label of every vertex; minimized in the objective~\eqref{eq:ilp-obj}, it equals the span. Distance constraint~\eqref{eq:ilp-radio} enforces the radio condition $|f(u) - f(v)| \ge \Delta_{uv}$ directly through conflict inequalities: whenever $x_{u,a}=1$ (i.e.\ $f(u)=a$), every label $b$ within the forbidden interval $|a-b| < \Delta_{uv}$ is excluded for $v$, since assigning both would force $x_{u,a}+x_{v,b}=2$. This avoids the big-$M$ linearization and the auxiliary direction variables entirely, as the disjunction $f(u)>f(v) \lor f(v)>f(u)$ is absorbed into the symmetric enumeration of all conflicting label pairs $(a,b)$. The same set of conflict pairs underlies the direct SAT encoding of Section~\ref{sec:SAT}.
	
	The model contains $n\,\mathrm{UB} + 1$ variables and $O(n^2\,\mathrm{UB}\,k)$ constraints, the latter dominated by the pairwise radio conflicts~\eqref{eq:ilp-radio}: each pair $(u,v)$ contributes $O(\mathrm{UB}\,\Delta_{uv})$ inequalities and $\Delta_{uv} \le k+1$. Although exact, the large number of conflict inequalities causes the search space to grow rapidly, limiting scalability on larger graphs. The formulation shares the assignment variables $x_{v,l}$ with the SAT encodings of
	Section~\ref{sec:SAT}, so the formulations admit the same set of feasible labelings over the domain $[1,\mathrm{UB}]$.

	\section{Related Works}
	\label{subsec:relatedworks}

	\begin{table}[htbp]
		\centering
		\caption{Summary of related works.}
		\label{tab:related-works}
		\setlength{\tabcolsep}{4pt}
		\renewcommand{\arraystretch}{1.2}
		\begin{tabular}{p{1.5cm} p{3cm} p{5.5cm} p{2.5cm}}
			\toprule
			\textbf{Reference} & \textbf{Graph families} & \textbf{Main contribution} & \textbf{Method} \\
			\midrule
			
			\multicolumn{4}{l}{\textit{Exact Theoretical Constructions}} \\
			\midrule
			
			\cite{LiuZhu2005}
			& Paths $P_n$, cycles $C_n$
			& Closed-form exact $rn(P_n)$, $rn(C_n)$
			& Vertex-ordering \\
			
			\cite{LiuXie2004}, \cite{LiuXie2009}
			& $C_n^2$, $P_n^2$
			& Exact $rn$ for squares of paths and cycles
			& Vertex-ordering \\
			
			\cite{Liu2008}
			& Trees
			& Exact $rn$(tree)
			& Vertex-ordering \\
			
			\cite{LiMakZhou2010}
			& Complete $m$-ary trees
			& Exact $rn$ and feasible labeling
			& Vertex-ordering \\
			
			\cite{Ortiz2011}
			& Generalized prism graphs
			& Exact $rn$ and feasible labeling
			& Vertex-ordering \\
			
			\cite{KolaPanigrahi2009a}, \cite{JuanLiu2012}
			& Paths $P_n$, cycles $C_n$
			& Exact antipodal number
			& Vertex-ordering \\
			
			\cite{KimSong2015}
			& Cartesian product $P_n \square K_m$
			& Exact $rn$ and feasible labeling
			& Vertex-ordering \\
			
			\cite{BantvaLiu2021}
			& Block graphs, line graphs of trees
			& Exact $rn$ and feasible labeling
			& N\&S condition \\
			
			\cite{Saha2022}
			& Full $m$-ary trees
			& Exact $rn$ and feasible labeling
			& Vertex-ordering \\
			
			\cite{VasoyaBantva2023}
			& Generalized Petersen $\square$ tree
			& Exact $rn$ and feasible labeling
			& N\&S condition \\
			
			\cite{VasoyaBantva2024}
			& Cartesian product (tree $\square\, K_m$)
			& Exact $rn$ and feasible labeling
			& N\&S condition \\
			
			\cite{Chakraborty2024}
			& Powers of paths $P_n^m$
			& Exact $rn$, linear-time labeling
			& Modified DGNS + constructive \\
			
			\midrule
			\multicolumn{4}{l}{\textit{Analytical Bounds}} \\
			\midrule
			
			\cite{Kchikech2007}
			& Trees, infinite paths
			& Upper bound on $rn$(tree)
			& Analytical bound \\
			
			\cite{ReddyIyer2011}
			& Binomial trees, Fibonacci trees
			& Tightened upper bound
			& Constructive heuristic \\
			
			\cite{KolaPanigrahi2015}
			& Arbitrary graphs
			& General analytical lower bound
			& Analytical lower bound \\
			
			\cite{Das2017}
			& General graphs
			& DGNS-type lower bound technique
			& DGNS lower bound \\
			
			\midrule
			\multicolumn{4}{l}{\textit{Heuristic Algorithms}} \\
			\midrule
			
			\cite{saha2012graph}
			& General connected graphs
			& Upper bound ($\mathrm{Ub}_{1}$), feasible labeling
			& Greedy \\
			
			\cite{badr2020upper}
			& General graph 
			& Improved upper bound ($\mathrm{Ub}_{3}$);
			& Iterated greedy \\
			
			\midrule
			\multicolumn{4}{l}{\textit{Exact Optimization}} \\
			\midrule
			
			\cite{badr2020upper}
			& General graph 
			& ILP formulation, exact in principle; limited scalability
			& ILP (LINGO solver) \\
			
			\midrule
			\textbf{This work}
			& General graph
			& Certified exact $rn(G)$ on solved instances; on growing-diameter families, tighter feasible labeling than $\mathrm{Ub}_{1}$, $\mathrm{Ub}_{3}$, and ILP; on bounded-diameter families, ILP CPLEX is preferable and the two methods are complementary.
			& Order-encoding SAT, incremental SAT, symmetry breaking \\
			\bottomrule
		\end{tabular}
	\end{table}

	The radio $k$-labeling problem originates from the channel assignment problem introduced by Hale~\cite{Hale1980}, in which frequencies must be assigned to transmitters so that graph distances model the interference constraints between communication devices. This line of research evolved into distance-constrained graph labeling problems, including the well-known $L(2,1)$-labeling of Griggs and Yeh~\cite{GriggsYeh1992}, who established the NP-completeness of the problem for general graphs, and was generalized by Chartrand et al.~\cite{Chartrand2001} through the notions of radio labeling and the radio $k$-chromatic number $\mathrm{rc}_k(G)$. Since then, the problem has attracted extensive attention from both the graph theory and the combinatorial optimization communities, and the resulting literature, summarized in Table~\ref{tab:related-works}, can be organized into four broad directions: exact theoretical constructions, analytical bounds, heuristic algorithms, and exact optimization formulations.
	
	The first and largest direction derives exact radio numbers for highly structured graph families through constructive combinatorial arguments, most often via carefully designed vertex orderings. Liu and Zhu~\cite{LiuZhu2005} introduced the vertex-ordering technique, constructing an ordering $(x_1,\ldots,x_n)$ and assigning labels recursively through $f(x_{i+1})-f(x_i)=k+1-d(x_i,x_{i+1})$, which under suitable distance conditions satisfies all pairwise radio constraints; this yielded the classical closed-form values $rn(P_n)$ and $rn(C_n)$. The methodology was subsequently applied to powers of paths and cycles~\cite{LiuXie2004,LiuXie2009}, antipodal labelings~\cite{KhennoufahTogni2005,JuanLiu2012,KolaPanigrahi2009a}, hypercubes~\cite{KhennoufahTogni2011}, generalized prism graphs~\cite{Ortiz2011}, trees~\cite{Liu2008}, complete $m$-ary trees~\cite{LiMakZhou2010}, Cartesian product graphs~\cite{KimSong2015}, and full $m$-ary trees~\cite{Saha2022}. A complementary line replaces explicit orderings by necessary-and-sufficient optimality conditions: Bantva~\cite{bantva2019} characterized optimal labelings through distance-sum maximization with vanishing slack, an approach later extended to block graphs and line graphs of trees~\cite{BantvaLiu2021}, to the Cartesian product of the generalized Petersen graph with trees~\cite{VasoyaBantva2023}, and to the product of a tree with a complete graph $K_m$~\cite{VasoyaBantva2024}. Other structural reductions include Sarkar and Adhikari~\cite{SarkarAdhikari2015}, who reduced radio labeling to a path-covering problem on graph powers for triangle-free and circulant graphs, and Chakraborty et al.~\cite{Chakraborty2024}, who obtained exact radio numbers for powers of paths via a modified DGNS framework with linear-time labeling. While elegant, these results depend strongly on graph symmetries or family-specific structure and do not naturally generalize.
	
	The second direction provides analytical lower and upper bounds for families where exact values remain unknown. On the upper-bound side, Kchikech et al.~\cite{Kchikech2007} bounded $rn$ for trees that are neither stars nor paths, and Reddy and Iyer~\cite{ReddyIyer2011} refined these estimates for binomial and binary Fibonacci trees, proving $rn(BT_k)\le 2(2^{k-1}-1)(k-1)+3\cdot2^{k-3}$. On the lower-bound side, Kola and Panigrahi~\cite{KolaPanigrahi2015} introduced a triameter-based framework, Saha and Panigrahi~\cite{SahaPanigrahi2015} a distance-matrix decomposition $\mathrm{rc}_k(G) = (n-1)(k+1) - \sum d(x_i,x_{i-1}) + \sum \varepsilon_i + 1$ with slack $\varepsilon_i\ge0$, and Das et al.~\cite{Das2017} the DGNS layered-decomposition bound, which strictly improves~\cite{KolaPanigrahi2015} on several families. Such bounds illuminate the asymptotic behavior of $rn$ but generally do not certify optimal labelings for finite instances.
	
	The third direction develops constructive heuristics. Saha and Panigrahi~\cite{saha2012graph} proposed an $O(n^3)$ greedy algorithm, denoted $\mathrm{Ub}_{1}$, that fixes the root label and assigns each vertex its minimum feasible label; it matches $rn(C_n)$ for all even $n$ up to $400$. Badr and Moussa~\cite{badr2020upper} iterated this construction over all roots to obtain the tighter bound $\mathrm{Ub}_{3}$ in $O(n^4)$ time, improving $\mathrm{Ub}_{1}$ on all tested families and also giving an exact $O(n)$ algorithm for $P_n$. Family-specific refinements followed for triangular and double triangular snakes~\cite{ELrokh2022} and for generalized friendship graphs~\cite{Alkasasbeh2023}. These methods are effective in practice but produce only feasible labelings and cannot certify optimality.
	
	The fourth and least-explored direction concerns exact optimization. Badr and Moussa~\cite{badr2020upper} formulated radio $k$-labeling as an integer linear program that is exact in principle but scales poorly beyond $n>100$ (the formulation and its limitations are detailed in Section~\ref{subsec:ilp}). More fundamentally, this and all existing exact approaches solve closely related optimization subproblems independently, discarding the information learned in previous iterations and thereby incurring substantial redundant search effort.
	
	Taken together, these limitations motivate the present work. Exact theoretical results remain confined to structured families; analytical bounds rarely yield certified labelings; heuristics provide no optimality guarantee; and the single existing exact formulation scales poorly and reuses no search information. To address these gaps, we introduce an exact incremental SAT-based framework that combines a compact order encoding with clause-learning SAT technology. Unlike heuristic methods, it certifies optimality whenever an UNSAT proof is obtained; unlike the existing ILP formulation, it reuses learned clauses across optimization iterations, substantially reducing redundant search effort and improving feasible solution quality on challenging instances.

	\section{SAT-based Approaches for Radio \texorpdfstring{$k$}{k}-labeling Problem}
	\label{sec:SAT}
	
	This section develops two Boolean satisfiability encodings of the radio $k$-labeling problem. Both reuse the modeling conventions of the ILP formulation in Section~\ref{subsec:ilp}, the label domain is fixed to $\{1, 2, \dots, \mathrm{UB}\}$, where $\mathrm{UB}$ is a precomputed upper bound on the span; for each ordered pair $(u,v)$ with $u<v$ the required separation is $\Delta_{uv} = k + 1 - d(u,v)$; and, as in constraint~\eqref{eq:ilp-assign}, every vertex receives exactly one label.

    
	
	\subsection{Direct Encoding}
	\label{subsec:direct-encoding}

	The direct encoding uses exactly the assignment variables of the ILP model, for every vertex $v \in V$ and label $l \in \{1,\dots,\mathrm{UB}\}$,
	$$
	x_{v,l} = 1 \iff f(v) = l,
	$$
	giving $n\,\mathrm{UB}$ Boolean variables, the Boolean counterpart of the ILP variables in~\eqref{eq:ilp-dom}. The radio condition is imposed purely through clausal constraints, organized into two groups.
	
	The clausal form of the assignment constraint~\eqref{eq:ilp-assign} consists of two parts. The \emph{at-least-one} clause, a single disjunction which guarantees $v$ is assigned some label, and a family of binary clauses for every pair $1\le a<b\le \mathrm{UB}$, which forbids $v$ from receiving two labels simultaneously,
	\begin{equation}
		\bigvee_{l=1}^{\mathrm{UB}} x_{v,l},
		\qquad
		\neg x_{v,a} \lor \neg x_{v,b}.
		\label{eq:d1}
	\end{equation}
	Together these two conditions enforce that each vertex receives exactly one label.
    The naive pairwise form in~\eqref{eq:d1} alone contributes $n\binom{\mathrm{UB}}{2}=O(n\,\mathrm{UB}^2)$ clauses.

	Mirroring the ILP conflict inequalities~\eqref{eq:ilp-radio}, the condition $|f(u)-f(v)|\geq\Delta_{uv}$ is enforced by forbidding, for each pair $(u,v)$ with $u<v$ and $\Delta_{uv}>0$, every violating label combination: for all $a,b\in\{1,\dots,\mathrm{UB}\}$ with $|a-b|<\Delta_{uv}$,
	\begin{equation}
		\neg x_{u,a} \lor \neg x_{v,b}.
		\label{eq:d2-conflict}
	\end{equation}
	Each pair $(u,v)$ contributes
	$$
	\bigl|\{(a,b):|a-b|<\Delta_{uv}\}\bigr| = \mathrm{UB} + 2\sum_{t=1}^{\Delta_{uv}-1}(\mathrm{UB}-t) = O(\mathrm{UB}\,\Delta_{uv}),
	$$
	and since $\Delta_{uv}\leq k+1$, the radio constraints~\eqref{eq:d2-conflict} contribute $O(n^2\,\mathrm{UB}\,k)$ clauses in the worst case.
	
    The direct encoding needs no auxiliary direction variables the disjunctive case $f(u)>f(v)$ or $f(v)>f(u)$ is absorbed into~\eqref{eq:d2-conflict}, but it has two structural drawbacks. First, $f(v)\geq c$ is not a single literal but the disjunction $\bigvee_{l\geq c}x_{v,l}$, so the radio condition must enumerate forbidden \emph{pairs} individually, an extra factor of $O(k)$ relative to the order encoding below. Second, propagation under unit resolution is weak, fixing $x_{u,a}=1$ excludes each conflicting label of $v$ only through a separate binary clause. 
    These two structural drawbacks motivate the order encoding below.
	
	\subsection{Order Encoding}
	\label{subsec:order-encoding}
	
	The order encoding represents each label $f(v)$ by a monotone chain of threshold variables together with the same mask variables $x_{v,l}$ used by the direct encoding and the ILP. For every vertex $v \in V$ and every label $l \in \{1, \dots, \mathrm{UB}\}$ we define the \emph{order} variables and the \emph{mask} variables
	$$
	g_{v,l} = 1 \iff f(v) \geq l,
	\qquad
	x_{v,l} = 1 \iff f(v) = l.
	$$
	The mask variables $x_{v,l}$ coincide exactly with the ILP assignment variables~\eqref{eq:ilp-dom}, so any satisfying assignment of the SAT model induces a feasible ILP solution and vice versa. The advantage of the threshold variables is twofold: (i) an inequality of the form $f(v) \geq c$ becomes the single literal $g_{v,c}$, and (ii) the encoding admits compact propagation of the arithmetic radio condition
	$|f(u) - f(v)| \geq \Delta_{uv}$.

	The constraints are organized into three groups in conjunctive normal form.
	The first group (O1) enforces domain consistency for the threshold encoding.
	Since $f(v) \geq l+1$ implies $f(v) \geq l$, the threshold variables are monotonically non-increasing in $l$, and in addition every vertex receives a label of at least $1$, so for every $v \in V$ and every
	$l \in \{1, \dots, \mathrm{UB}-1\}$,
	\begin{equation}
		(\neg g_{v,l+1} \lor g_{v,l}),
		\qquad
		g_{v,1}.
		\label{eq:o1}
	\end{equation}
	Together the clauses in~\eqref{eq:o1} guarantee that each vertex is assigned exactly one label, the clausal counterpart of the assignment constraint~\eqref{eq:ilp-assign}. Besides, they enforce the at-most-one condition implicitly through monotonicity rather than through an explicit summation or the $O(n\,\mathrm{UB}^2)$ at-most-one clauses in~\eqref{eq:d1} of the direct encoding.
	
	The second group (O2) links the threshold and mask variables. The variable $x_{v,l}$ holds if and only if $f(v) \geq l$ and $f(v) \not\geq l+1$, so for every $v \in V$ and every $l \in \{1, \dots, \mathrm{UB}-1\}$ this equivalence is captured by the three clauses
	\begin{equation}
		x_{v,l} \rightarrow g_{v,l}, \qquad
		x_{v,l} \rightarrow \neg g_{v,l+1}, \qquad
		(g_{v,l} \land \neg g_{v,l+1}) \rightarrow x_{v,l}.
		\label{eq:o2}
	\end{equation}
	For the boundary case $l = \mathrm{UB}$, since $g_{v,\mathrm{UB}+1}$ does not exist, the equivalence reduces to
	$x_{v,\mathrm{UB}} \leftrightarrow g_{v,\mathrm{UB}}$, encoded by
	\begin{equation}
		\neg x_{v,\mathrm{UB}} \lor g_{v,\mathrm{UB}}, \qquad
		\neg g_{v,\mathrm{UB}} \lor x_{v,\mathrm{UB}}.
		\label{eq:o2-boundary}
	\end{equation}
	
	The third group (O3) is the radio distance constraint, the clausal counterpart of~\eqref{eq:ilp-radio}. For each pair $(u, v)$ with $u < v$ and $\Delta_{uv} > 0$, whenever $x_{u,a}=1$ (i.e., $f(u) = a$) the label of $v$ must lie outside the forbidden interval $(a - \Delta_{uv},\, a + \Delta_{uv})$,
	$$
	x_{u,a} \implies \big(f(v) \leq a - \Delta_{uv}\big) \,\lor\,
	\big(f(v) \geq a + \Delta_{uv}\big).
	$$
	Using the threshold variables, the two disjuncts translate to single literals
	$$ f(v) \leq a - \Delta_{uv} \Leftrightarrow \neg g_{v,\, a - \Delta_{uv} + 1},
	\qquad
	f(v) \geq a + \Delta_{uv} \Leftrightarrow g_{v,\, a + \Delta_{uv}}. $$
	Combined with the antecedent $\neg x_{u,a}$, this yields a \emph{single} clause per triple $(u, v, a)$ in contrast to the $O(\mathrm{UB}\,\Delta_{uv})$ binary clauses~\eqref{eq:d2-conflict} the direct encoding needs for the same pair. The indices $a - \Delta_{uv} + 1$ and $a + \Delta_{uv}$ may fall outside $[1, \mathrm{UB}]$, in which case the corresponding literal is dropped, giving three cases: in Case~1 both disjuncts are feasible ($a > \Delta_{uv} - 1$ and $a + \Delta_{uv} \leq \mathrm{UB}$); in Case~2 only the lower disjunct is feasible ($a + \Delta_{uv} > \mathrm{UB}$), so $f(v)$ is pushed below $a$; and in Case~3 only the upper disjunct is feasible ($a \leq \Delta_{uv} - 1$), so $f(v)$ is pushed above $a$,
	\begin{equation}
		\underbrace{\neg x_{u,a} \lor \neg g_{v, a - \Delta_{uv} + 1} \lor g_{v,\, a + \Delta_{uv}}}_{\text{Case 1}},
		\quad
		\underbrace{\neg x_{u,a} \lor \neg g_{v, a - \Delta_{uv} + 1}}_{\text{Case 2}},
		\quad
		\underbrace{\neg x_{u,a} \,\lor\, g_{v,\, a + \Delta_{uv}}}_{\text{Case 3}}.
		\label{eq:o3}
	\end{equation}
	
	The order encoding is correct in the sense that the satisfying assignments of the CNF formula $\Phi$ formed by the clause groups \textnormal{(O1)}--\textnormal{(O3)} are in one-to-one correspondence with the radio labelings of $G$ over the domain $[1,\mathrm{UB}]$. Fix any satisfying assignment. By the unit clause $g_{v,1}$ in~\eqref{eq:o1} we have $g_{v,1}=1$, and by the monotonicity clauses in~\eqref{eq:o1} the truth values $g_{v,1}\ge g_{v,2}\ge\cdots\ge g_{v,\mathrm{UB}}$ form a non-increasing $0/1$ chain, so there is a unique index $f(v)=\max\{l:g_{v,l}=1\}\in[1,\mathrm{UB}]$; the linking clauses~\eqref{eq:o2}--\eqref{eq:o2-boundary} then force $x_{v,l}=1$ exactly for $l=f(v)$, so each vertex receives exactly one label, which is the clausal counterpart of the assignment constraint~\eqref{eq:ilp-assign}. For any pair $u<v$ with $\Delta_{uv}>0$, writing $a=f(u)$, the clause of group~\textnormal{(O3)} in~\eqref{eq:o3} for the triple $(u,v,a)$ forces $f(v)\le a-\Delta_{uv}$ or $f(v)\ge a+\Delta_{uv}$ through the threshold literals $\neg g_{v,\,a-\Delta_{uv}+1}$ and $g_{v,\,a+\Delta_{uv}}$ (literals outside $[1,\mathrm{UB}]$ are dropped precisely because the corresponding labels are infeasible), so $|f(u)-f(v)|\ge\Delta_{uv}=k+1-d(u,v)$ and $f$ is a valid radio labeling of span at most $\mathrm{UB}$. Conversely, given any radio labeling $f$ of span at most $\mathrm{UB}$, setting $g_{v,l}=1$ iff $f(v)\ge l$ and $x_{v,l}=1$ iff $f(v)=l$ satisfies \textnormal{(O1)} and \textnormal{(O2)} by construction and \textnormal{(O3)} because the radio condition places $f(v)$ outside the forbidden interval $(a-\Delta_{uv},a+\Delta_{uv})$ whenever $f(u)=a$. The two maps are mutually inverse, establishing the claimed correspondence.
	
	\subsection{Comparison of the Two Encodings}
	\label{subsec:encoding-comparison-discussion}
	
	Both SAT encodings represent the same feasibility problem as the ILP model of Section~\ref{subsec:ilp} over the identical label domain $[1, \mathrm{UB}]$ and share the mask variables $x_{v,l}$, so the three formulations are logically equivalent in the labelings they admit. They differ only in size and propagation behavior, summarized in Table~\ref{tab:encoding-comparison}.
	The order encoding doubles the variable count (adding the ordered variables $g_{v,l}$) but reduces the label-assignment clauses from $O(n\,\mathrm{UB}^2)$ in~\eqref{eq:d1} to $O(n\,\mathrm{UB})$ in~\eqref{eq:o1}, and the radio clauses from $O(n^2\,\mathrm{UB}\,k)$ in~\eqref{eq:d2-conflict} to $O(n^2\,\mathrm{UB})$ in~\eqref{eq:o3}, eliminating the dependence on $k$.

	\begin{table}[h]
		\centering
		\caption{Size of the two SAT encodings.}
		\label{tab:encoding-comparison}
		\begin{tabular}{lcc}
			\toprule
			\textbf{Component} & \textbf{Direct SAT} & \textbf{Order SAT} \\
			\midrule
			Boolean / binary vars & $n\,\mathrm{UB}$ & $2n\,\mathrm{UB}$ \\
			Assignment clauses    & $O(n\,\mathrm{UB}^2)$ & $O(n\,\mathrm{UB})$ \\
			Radio constraints     & $O(n^2\,\mathrm{UB}\,k)$ & $O(n^2\,\mathrm{UB})$ \\
			\bottomrule
		\end{tabular}
	\end{table}
	
	\subsection{Incremental SAT Solving Framework}
	\label{subsec:increSAT}

	The incremental optimization framework is independent of the underlying feasibility encoding and applies identically to both the direct and the order encodings of Sections~\ref{subsec:direct-encoding} and~\ref{subsec:order-encoding}. It follows an \emph{incremental SAT} paradigm that a single solver instance is built once and progressively strengthened across iterations, in three phases.
	
	In phase 1 (preprocessing), we compute the all-pairs shortest-path matrix and set $k=\mathrm{diam}(G)$. An initial upper bound $\mathrm{UB}$ comes from a graph-specific heuristic, and the lower bound is set to $1$.
    
	The chosen encoding is invoked with parameter $m=\mathrm{UB}$ in phase 2, producing a CNF formula and initializing an incremental SAT solver instance. The label domain is fixed for the entire run, no variables are added afterward. The value $\mathrm{UB} = n(k+1)-k$ is a valid domain bound. Ordering the vertices arbitrarily as $v_1,\ldots,v_n$ and assigning $f(v_i)=1+(i-1)(k+1)$ spaces labels by $k+1$, so for every $i<j$,
	$$
	|f(v_i)-f(v_j)| = (j-i)(k+1) \geq k+1 > k \geq k+1-d(v_i,v_j),
	$$
	since $d(v_i,v_j)\geq1$. This assignment is therefore always feasible, with maximum label $f(v_n)=n(k+1)-k$, so $rn(G)\leq n(k+1)-k$ and the domain $[1,\mathrm{UB}]$ is guaranteed to contain an optimal assignment.
    
	In phase 3, the algorithm repeatedly re-solves the incrementally strengthened formula. Each \textsc{Sat} result tightens the span and the search continues with a stricter bound by \textsc{ForbidSpansAtOrAbove}$(\beta, \mathtt{solver})$. The loop terminates either when the solver returns \textsc{Unsat} for the current bound, certifying that the span is optimal, or when the time limit is reached, in which case the span is returned as a feasible but uncertified bound.

\begin{algorithm}
		\caption{Incremental SAT for Radio $k$-Labeling}
		\label{alg:incremental-sat}
		\begin{algorithmic}[1]
			\Require Graph $G=(V,E)$, $|V|=n$; initial upper bound $\mathtt{ub}$; time limit $T$
			\Ensure Best span $\beta$ found within $T$; certified optimal if terminated via \textsc{Unsat}
			
			\Statex \textit{Phase 1: Preprocessing}
			\State $\mathrm{dist} \gets$ all-pairs distances via BFS
			\State $k \gets \mathrm{diam}(G)$
			
			\Statex \textit{Phase 2: Build SAT model once}
			\State $\mathtt{solver} \gets \textsc{BuildSAT}(n, \mathtt{ub}, \mathrm{dist}, k)$
			\Comment{domain $[1,\mathtt{ub}]$, fixed}
			\State $\beta \gets \mathtt{ub}$
			\Comment{feasible by construction, Section~\ref{subsec:increSAT}}
			
			\Statex \textit{Phase 3: Incrementally tighten the span}
			\While{time remains}
			\State $(\mathtt{status}, \mathtt{model}) \gets \textsc{Solve}(\mathtt{solver}, T_{\mathrm{remain}})$
			
			\If{$\mathtt{status} = \textsc{Timeout}$}
			\State \Return $\beta$ \Comment{Time-Out: last confirmed feasible span}
			\ElsIf{$\mathtt{status} = \textsc{Sat}$}
			\State $\beta \gets \textsc{ExtractSpan}(\mathtt{model})$ \Comment{$\max_v f(v)$}
			\State \textsc{ForbidSpansAtOrAbove}$(\beta, \mathtt{solver})$
			\Comment{next \textsc{Solve} tests span $\le \beta-1$}
			\Else
			\State \Return $\beta$ \Comment{\textsc{Unsat}: no span $<\beta$ exists, so $\beta$ is optimal}
			\EndIf
			\EndWhile
			\State \Return $\beta$
		\end{algorithmic}
	\end{algorithm}
	
			
	
    All conflict clauses learned by SAT solver in earlier iterations remain valid and continue pruning the search in later ones, the central source of savings over independent SAT calls. Rather than decreasing admissible span bound by one unit per iteration, we use the actual span $\beta$ extracted from the satisfying model, which is often much smaller than the naively decremented bound. This lets the search skip trivially feasible values and reduces the number of solver calls.  
	
	
	\subsection{Symmetry Breaking}
	\label{subsec:symbreak}
	
	For cycle graphs $C_n$, the graph structure exhibits rotational symmetry. If a labeling is feasible, then rotating the labels around the cycle produces another equivalent feasible solution. Consequently, the SAT solver may spend considerable time exploring many rotationally equivalent assignments.
	To eliminate this symmetry, one vertex is fixed to a predefined label. Without loss of generality, the first vertex is assigned the smallest label
	$f(0) = 1$.
	This constraint removes equivalent rotational solutions and significantly reduces the number of symmetric assignments explored during SAT solving.

	Fixing $f(0)=1$ never discards an optimal solution, that is, $C_n$ admits a radio labeling of span $s$ if and only if it admits one of span $s$ with $f(0)=1$. To see this, let $f$ be any radio labeling of span $s$. Subtracting $\min_v f(v)-1$ from every label is a translation that leaves all differences $|f(u)-f(v)|$, hence all radio constraints, unchanged and keeps the span equal to $s$, so we may assume the smallest label equals $1$. Let $w$ be a vertex with $f(w)=1$. Since $C_n$ is vertex-transitive, its rotational automorphism group acts transitively on $V$, so there is a rotation $\rho$ with $\rho(0)=w$; as $\rho$ preserves all pairwise distances, $f\circ\rho$ is again a radio labeling of span $s$ and satisfies $(f\circ\rho)(0)=f(w)=1$. Hence an optimal labeling with $f(0)=1$ always exists, and the constraint removes only redundant rotated copies.

	\section{Experiments and Results}
	\label{sec:exp&result}
	
	\subsection{Experimental Setup}
	\label{subsec:setup}
	
	All experiments were conducted on a personal computer equipped with an Intel(R) Core(TM) i7-10870H CPU running at 2.20 GHz and 24 GB of RAM.
    The Direct Encoding (Direct-SAT) and Order Encoding (Order-SAT) were solved using the CaDiCaL 1.9.5 SAT solver. For Direct-SAT, we evaluated three encodings for the exactly-one assignment constraints in~\eqref{eq:d1}: the pairwise encoding, and two cardinality encodings provided by PySAT \cite{ignatiev2018pysat}, namely Sequential Counter (seqcounter) and Cardinality Network (cardnetwrk). This comparison allows us to assess the impact of different encodings for exactly-one constraints on the performance of the SAT solver. 
    The ILP model~\eqref{eq:ilp-obj} was solved using two commercial optimization solvers, IBM ILOG CPLEX (Version 22.2.0) and Gurobi (Version 13.0.0), denoted ILP CPLEX and ILP Gurobi, respectively. All methods were evaluated under the same time limit of 1,200 seconds.
	
	The experimental evaluation was conducted on the benchmark dataset used in Ub3 \cite{badr2020upper}. The benchmark instances were generated with increasing numbers of vertices in order to evaluate scalability and optimization capability under different graph sizes.
	The benchmark comprises $146$ instances from nine graph families: paths $P_n$ ($n=1$--$25$), cycles $C_n$ ($n=3$--$25$), triangular snakes $\Delta_k$ ($k=1$--$15$), $kC_4$-snakes ($k=1$--$15$), $kC_6$-snakes ($k=1$--$15$), ladder graphs $L_k$ ($k=1$--$15$), book graphs $B_k$ ($k=1$--$15$), friendship graphs $F_k$ ($k=1$--$15$), and binomial trees $BT_k$ ($k=0$--$7$).
    
    The overall performance of each method is summarized using five metrics. $\#Best$ denotes the number of instances on which a method achieves the best bound among all compared methods. $\#Opt$ is the number of instances for which optimality is certified. $\#Solved$ reports the number of instances solved within the time limit by finding a feasible solution. $\#New$ counts the instances for which a method improves upon the previously best-known upper bound. \textit{Avg. Rank} is the average rank obtained from the Friedman test, with lower values indicating better overall performance.

\begin{figure}[H]
		\centering
		\begin{subfigure}[b]{0.49\textwidth}
			\centering
			\includegraphics[width=\textwidth]{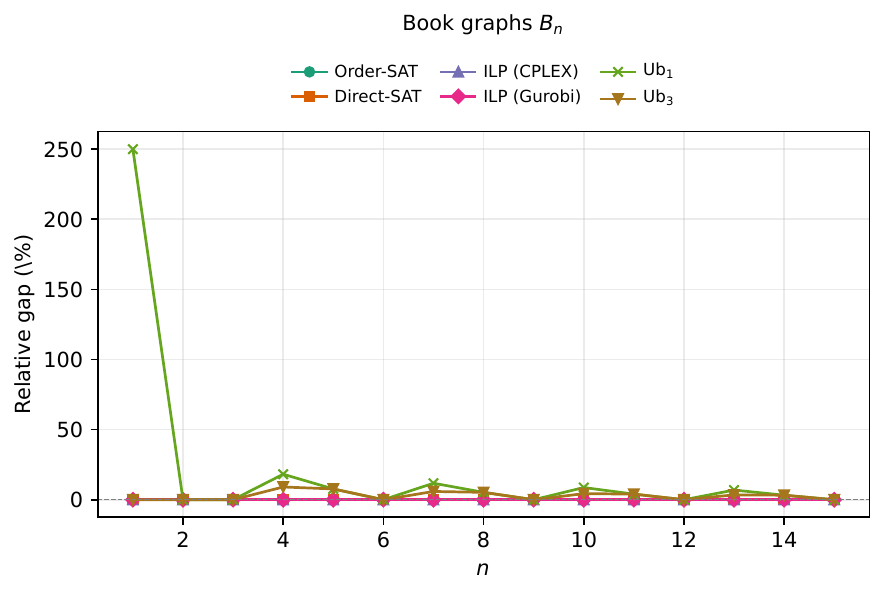}
			\caption{Book graphs $B_n$}
			\label{fig:gap-book}
		\end{subfigure}
		\hfill
		\begin{subfigure}[b]{0.49\textwidth}
			\centering
			\includegraphics[width=\textwidth]{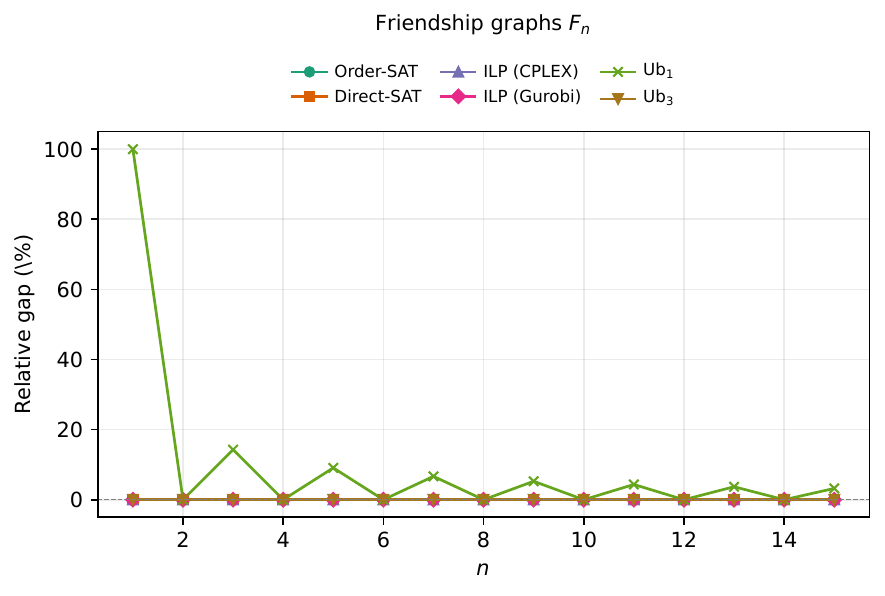}
			\caption{Friendship graphs $F_n$}
			\label{fig:gap-friendship}
		\end{subfigure}
		\caption{Relative gap, bounded-diameter families.}
		\label{fig:gap-bounded}
	\end{figure}

\begin{figure}[H]
		\centering
		\begin{subfigure}[b]{0.48\textwidth}
			\centering
			\includegraphics[width=\textwidth]{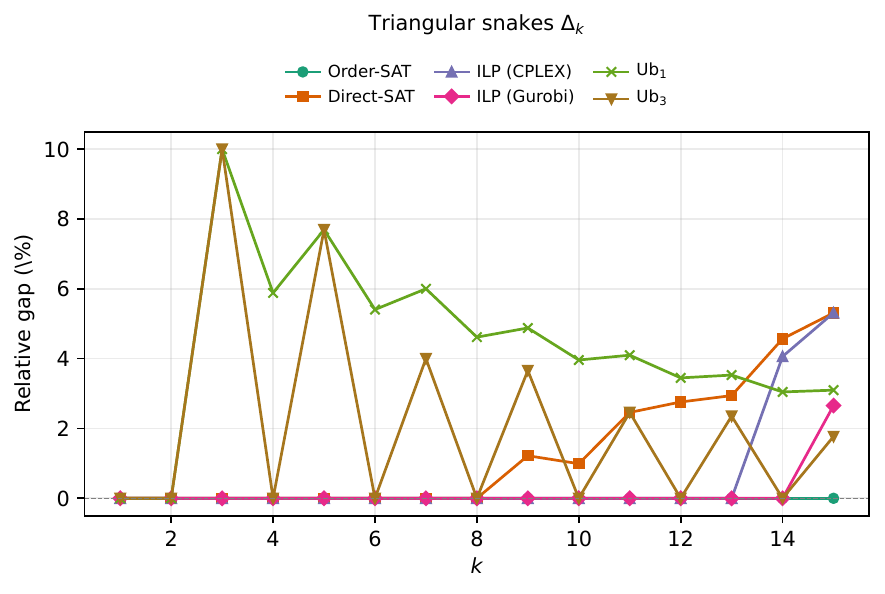}
			\caption{Triangular snakes $\Delta_k$}
			\label{fig:gap-trisnake}
		\end{subfigure}
		\hfill
		\begin{subfigure}[b]{0.48\textwidth}
			\centering
			\includegraphics[width=\textwidth]{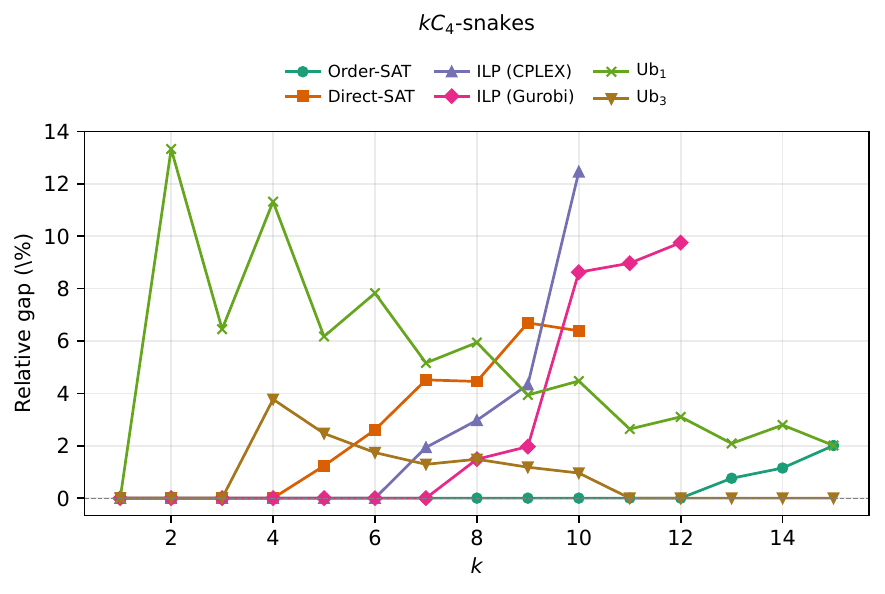}
			\caption{$kC_4$-snakes}
			\label{fig:gap-c4snake}
		\end{subfigure}
		
		\vspace{1em}
		\begin{subfigure}[b]{0.48\textwidth}
			\centering
			\includegraphics[width=\textwidth]{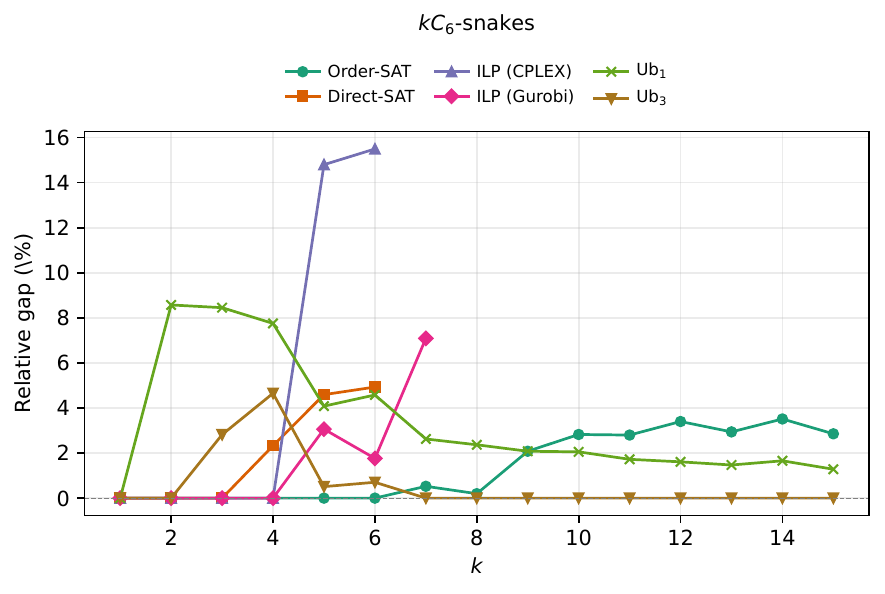}
			\caption{$kC_6$-snakes}
			\label{fig:gap-c6snake}
		\end{subfigure}
		\hfill
		\begin{subfigure}[b]{0.48\textwidth}
			\centering
			\includegraphics[width=\textwidth]{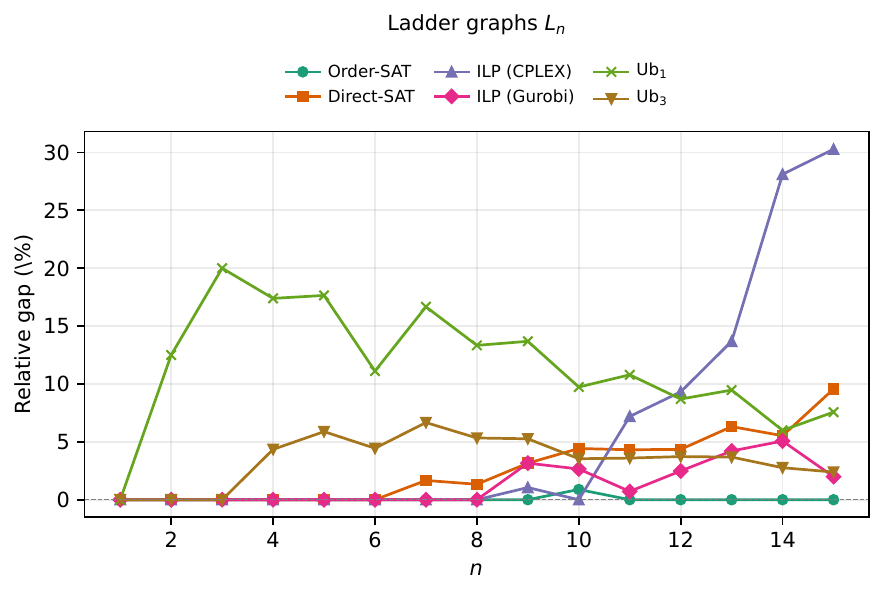}
			\caption{Ladder graphs $L_n$}
			\label{fig:gap-ladder}
		\end{subfigure}
		
		\vspace{1em}
		\begin{subfigure}[b]{0.6\textwidth}
			\centering
			\includegraphics[width=\textwidth]{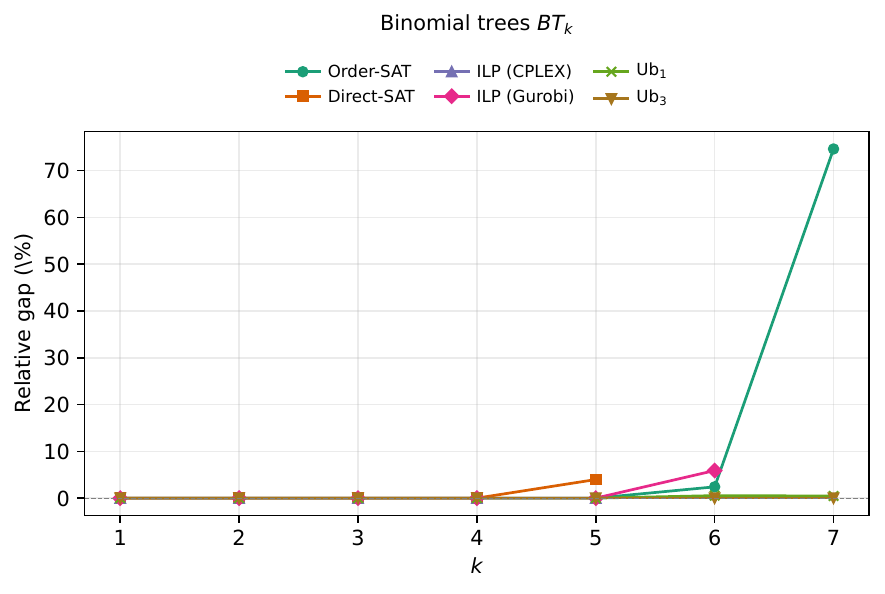}
			\caption{Binomial trees $BT_k$}
			\label{fig:gap-bitree}
		\end{subfigure}
		\caption{Relative gap, growing-diameter families.}
		\label{fig:gap-growing}
	\end{figure}

    Per-instance relative-gap plots, defined as $100\cdot(s-rn^\ast)/rn^\ast$ for a returned span $s$ against the best-known span $rn^\ast$, with $0\%$ marking the best result, are presented alongside their respective discussions, Figure~\ref{fig:gap-pathcycle} in Section~\ref{subsec:cycpath} for paths and cycles, and Figures~\ref{fig:gap-growing} and~\ref{fig:gap-bounded} in Section~\ref{subsec:other-families} for the remaining families.

	\subsection{Computational results for cycle and path graphs}\label{subsec:cycpath}

    For paths and cycles, the exact radio number $\mathrm{Ub}_0$ has been established in closed form by Liu and Zhu~\cite{LiuZhu2005}. This optimal value serves as the reference for evaluating the upper bounds $\mathrm{Ub}_1$~\cite{saha2012graph}, $\mathrm{Ub}_3$~\cite{badr2020upper}, as well as the SAT and ILP methods (Table~\ref{tab:friedman-pathcycle}). The $\mathrm{Ub}_0$ values were recomputed directly from the closed-form formulas given in the original paper, since this produces more accurate results than those reported in $\mathrm{Ub}_3$~\cite{badr2020upper}. In contrast, the $\mathrm{Ub}_1$ and $\mathrm{Ub}_3$ upper bounds were generated by our own implementations of the constructive algorithms described in their respective original publications, rather than by reproducing the reported results.
    

	\begin{figure}[H]
		\centering
		\begin{subfigure}[b]{0.48\textwidth}
			\centering
			\includegraphics[width=\textwidth]{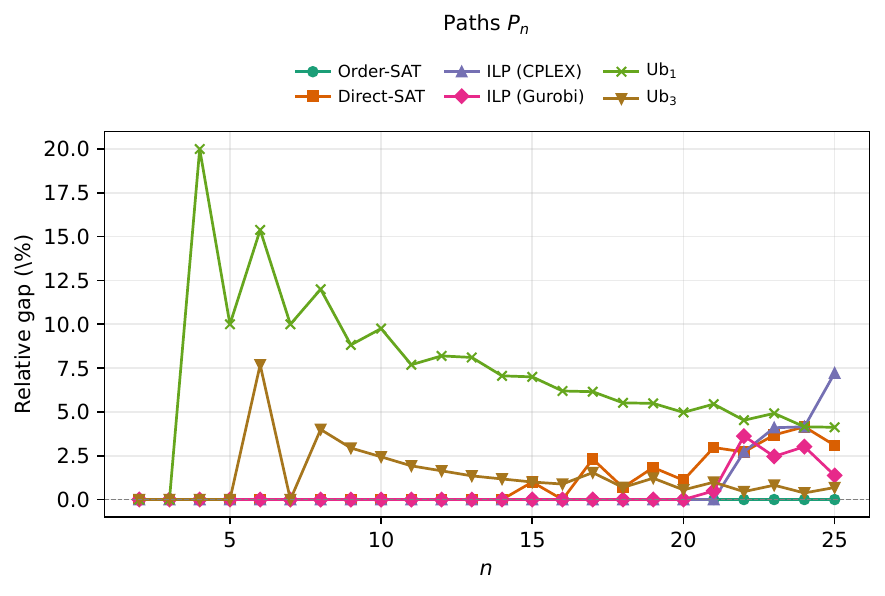}
			\caption{Paths $P_n$}
			\label{fig:gap-path}
		\end{subfigure}
		\hfill
		\begin{subfigure}[b]{0.48\textwidth}
			\centering
			\includegraphics[width=\textwidth]{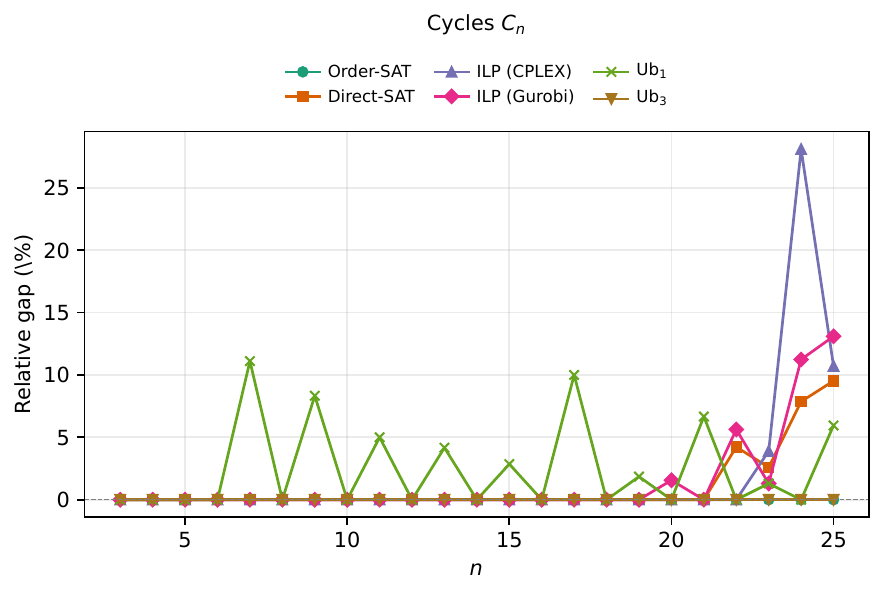}
			\caption{Cycles $C_n$}
			\label{fig:gap-cycle}
		\end{subfigure}
		\caption{Relative gaps on the path and cycle graph families.}
		\label{fig:gap-pathcycle}
	\end{figure}
	
    Figure~\ref{fig:gap-pathcycle} illustrates the relative gap of all methods on the path and cycle benchmark instances, measured with respect to the exact radio number $\mathrm{Ub}_0$. Since $\mathrm{Ub}_0$ represents the proven optimum for these graph families, the relative gap directly reflects the quality of the best values produced by each method. Lower relative gaps indicate solutions closer to the optimum, enabling a clear visual comparison of the optimization performance of the heuristic, SAT-based, and ILP-based approaches.

    \begin{table}[htbp]
		\centering
		\caption{Performance summary of the compared methods on path and cycle graphs.}
		\label{tab:friedman-pathcycle}
		\setlength{\tabcolsep}{3.5pt}
		\renewcommand{\arraystretch}{1.15}
		\begin{tabular}{l l ccccccc}
			\toprule
			Family & Metric & Order-SAT & Direct-SAT & ILP & ILP & $\mathrm{Ub}_{0}$ & $\mathrm{Ub}_{1}$ & $\mathrm{Ub}_{3}$  \\
			& &  & {\scriptsize (seqcounter)} & {\scriptsize CPLEX} & {\scriptsize Gurobi} & {\scriptsize\cite{LiuZhu2005}} & {\scriptsize\cite{saha2012graph}} & {\scriptsize\cite{badr2020upper}}  \\
			\midrule
			\multirow{3}{*}{Path $P_n$ (25)}
			& \#Best & 24 & 15 & 21 & 20 & \textbf{25} & 3 & 6  \\
			& \#Opt  & 15 & 11 & 19 & 20 & \textbf{25} & 0 & 0  \\
			& AVG. Rank   & 2.82 & 4.20 & 3.50 & 3.36 & \textbf{2.78} & 6.56 & 4.78  \\
			\midrule
			\multirow{3}{*}{Cycle $C_n$ (23)}
			& \#Best & \textbf{23} & 19 & 20 & 18 & \textbf{23} & 13 & \textbf{23}  \\
			& \#Opt  & 17 & 14 & 17 & 15 & \textbf{23} & 0 & 0  \\
			& AVG. Rank   & \textbf{3.50} & 4.07 & 4.11 & 4.33 & \textbf{3.50} & 4.91 & \textbf{3.50}  \\
			\bottomrule
		\end{tabular}
	\end{table}
    
	On path graphs, Order-SAT substantially improves upon the latest published heuristic $\mathrm{Ub}_3$, obtaining tighter upper bounds on $19$ of the $25$ benchmark instances. Compared with the two commercial ILP optimizers, CPLEX and Gurobi, Order-SAT consistently returns tighter feasible solutions on the more challenging instances, although the ILP solvers certify optimality for a larger number of cases. Moreover, Order-SAT clearly outperforms Direct-SAT, demonstrating the effectiveness of the proposed order encoding over the direct encoding formulation. 
    Order-SAT achieves the best mean rank among all practical methods (2.82), second only to the exact reference $\mathrm{Ub}_0$ (2.78), and ranks ahead of Gurobi, CPLEX, Direct-SAT, $\mathrm{Ub}_1$, and $\mathrm{Ub}_3$.
	
   On cycle graphs, Order-SAT matches the exact radio number $\mathrm{Ub}_0$ on every instance for which optimality is certified, proving optimality for $17$ of the $25$ benchmark instances. This represents the highest number of optimality certificates among the exact methods evaluated on this graph family. The effectiveness of Order-SAT is further enhanced by the proposed symmetry-breaking constraints, which substantially improve the solver's ability to certify optimality while preserving solution quality. Detailed results are reported in Tables~\ref{tab:compare-path-full} and~\ref{tab:compare-cycle-full}.

   Table~\ref{tab:cycle-fix-vertex} evaluates the proposed symmetry-breaking strategy, which fixes one vertex label ($f(0)=1$) to eliminate the rotational symmetry of cycle graphs (Section~\ref{subsec:symbreak}). Implemented as a unit clause on $g_{0,1}$, this constraint enables preprocessing to simplify the SAT instance, reducing the number of clauses by an average of $12.8\%$ ($7.4$--$21.8\%$). It also consistently accelerates solving, yielding speedups from  $1.47\times$ ($n{=}7$) to $24.35\times$ ($n{=}16$) on all jointly certified instances and enabling four additional optimality certificates ($C_{19}$, $C_{20}$, $C_{22}$, and $C_{25}$). Only $C_{24}$ exhibits a slightly weaker upper bound ($90$ vs.\ $89$), while $C_{23}$ and $C_{24}$ remain unsolved under both configurations. Overall, symmetry breaking simultaneously reduces the encoding size and search space, making it the preferred configuration for cycle graphs.
	

\begin{table}[ht!]
	\caption{Fix-one-vertex symmetry breaking on cycles $C_n$.}
	\label{tab:cycle-fix-vertex}
	\centering
	\tiny
	\setlength{\tabcolsep}{2.5pt}
	\renewcommand{\arraystretch}{1.25}
	\renewcommand{\tabularxcolumn}[1]{m{#1}}
	\begin{tabularx}{\textwidth}{>{\centering\arraybackslash}p{1cm} XXXXX XXXXX}
	\hline 
	\multirow{2}{*}{$n$} & \multicolumn{5}{c}{\textbf{Order-SAT (without SB)}} & \multicolumn{5}{c}{\textbf{Order-SAT (with SB)}} \\
	& \#Vars & \#Clauses & $rn(C_n)$ & Status & Time (s) & \#Vars & \#Clauses & $rn(C_n)$ & Status & Time (s) \\ \hline
	3  & 30 & {39} & 2 & OPT & 0.16 & 30 & \textbf{31} & 2 & OPT & \textbf{0.09} \\
	4  & 80 & {124} & 4 & OPT & 0.16 & 80 & \textbf{97} & 4 & OPT & \textbf{0.06} \\
	5  & 130 & {240} & 4 & OPT & 0.16 & 130 & \textbf{192} & 4 & OPT & \textbf{0.09} \\
	6  & 252 & {525} & 7 & OPT & 0.21 & 252 & \textbf{425} & 7 & OPT & \textbf{0.13} \\
	7  & 350 & {812} & 9 & OPT & 0.25 & 350 & \textbf{668} & 9 & OPT & \textbf{0.17} \\
	8  & 576 & {1\,480} & 13 & OPT & 0.38 & 576 & \textbf{1\,235} & 13 & OPT & \textbf{0.23} \\
	9  & 738 & {2\,106} & 12 & OPT & 0.48 & 738 & \textbf{1\,759} & 12 & OPT & \textbf{0.25} \\
	10  & 1\,100 & {3\,405} & 17 & OPT & 0.66 & 1\,100 & \textbf{2\,919} & 17 & OPT & \textbf{0.39} \\
	11  & 1\,342 & {4\,477} & 20 & OPT & 1.01 & 1\,342 & \textbf{3\,866} & 20 & OPT & \textbf{0.51} \\
	12  & 1\,872 & {6\,708} & 26 & OPT & 1.58 & 1\,872 & \textbf{5\,861} & 26 & OPT & \textbf{0.70} \\
	13  & 2\,210 & {8\,528} & 24 & OPT & 1.29 & 2\,210 & \textbf{7\,507} & 24 & OPT & \textbf{0.72} \\
	14  & 2\,940 & {12\,061} & 31 & OPT & 2.30 & 2\,940 & \textbf{10\,695} & 31 & OPT & \textbf{1.21} \\
	15  & 3\,390 & {14\,730} & 35 & OPT & 49.05 & 3\,390 & \textbf{13\,162} & 35 & OPT & \textbf{3.49} \\
	16  & 4\,352 & {19\,984} & 43 & OPT & 115.68 & 4\,352 & \textbf{17\,943} & 43 & OPT & \textbf{4.75} \\
	17  & 4\,930 & {23\,970} & 40 & OPT & 8.37 & 4\,930 & \textbf{21\,513} & 40 & OPT & \textbf{2.23} \\
	18  & 6\,156 & {31\,437} & 49 & OPT & 35.04 & 6\,156 & \textbf{28\,511} & 49 & OPT & \textbf{6.90} \\
	19  & 6\,878 & {36\,803} & 54 & TO & 1200.12 & 6\,878 & \textbf{33\,373} & \textbf{54} & \textbf{OPT} & \textbf{183.77} \\
	20  & 8\,400 & {47\,020} & 64 & TO & 1200.13 & 8\,400 & \textbf{42\,829} & \textbf{64} & \textbf{OPT} & \textbf{298.08} \\
	21  & 9\,282 & {54\,432} & 60 & OPT & 201.90 & 9\,282 & \textbf{49\,738} & 60 & OPT & \textbf{11.00} \\
	22  & 11\,132 & {68\,013} & 71 & TO & 1200.12 & 11\,132 & \textbf{62\,435} & \textbf{71} & \textbf{OPT} & \textbf{188.79} \\
	23  & 12\,190 & {77\,464} & 77 & TO & 1200.09 & 12\,190 & \textbf{71\,242} & 77 & TO & 1217.37 \\
	24  & 14\,400 & {95\,064} & \textbf{89} & TO & 1200.12 & 14\,400 & \textbf{87\,755} & 90 & TO & 1207.48 \\
	25  & 15\,650 & {107\,450} & 84 & TO & 1200.03 & 15\,650 & \textbf{99\,487} & \textbf{84} & \textbf{OPT} & \textbf{870.49} \\ \hline
	\end{tabularx}
	\vspace{2pt}
\end{table}

	\subsection{Computational results for other graph families}\label{subsec:other-families}

	The five remaining families span a spectrum of diameter behaviour, which the experiments identify as the single most predictive factor for the relative performance of the two exact methods. A \emph{$kC_m$-snake} chains $k$ copies of the cycle $C_m$ at shared cut vertices, so its diameter grows as $\lfloor m/2\rfloor\cdot k$; the \emph{ladder} $L_n = P_2 \square P_n$ has $2n$ vertices and diameter $n$; the \emph{book} $B_n = K_{1,n} \square K_2$ has $2n{+}2$ vertices but constant diameter $3$; the \emph{friendship} graph $F_n$ joins $n$ triangles at a common vertex and has constant diameter $2$; and the \emph{binomial tree} $BT_k$ has $2^k$ vertices and diameter $k$. The full per-instance numbers are reported in Appendix~\ref{app:detailed}; the discussion below is organized by the two regimes that emerge, namely families whose diameter grows with size and families whose diameter remains bounded.
	
	\begin{table}[ht!]
		\caption{Results by graph family.}
		\label{tab:summary}
		\tiny
		\centering
		\setlength{\tabcolsep}{2.5pt}
		\renewcommand{\arraystretch}{1.25}
		\renewcommand{\tabularxcolumn}[1]{m{#1}}
		\begin{tabularx}{\textwidth}{X l cccccc}
			\hline
			\textbf{Dataset} & \textbf{Metric} & \textbf{Order-SAT} & \textbf{Direct-SAT} & \textbf{ILP CPLEX} & \textbf{ILP Gurobi} & \textbf{$\mathrm{Ub}_{1}$}~\cite{saha2012graph} & \textbf{$\mathrm{Ub}_{3}$}~\cite{badr2020upper}  \\ \hline
			
			\multirow{5}{*}{Tri.\ snake $\Delta_k$ (15 inst.)}
			& \#Solved  & \textbf{15} & \textbf{15} & \textbf{15} & \textbf{15} & \textbf{15} & \textbf{15} \\
			& \#Best    & \textbf{15} & 8 & 13 & 14 & 2 & 8  \\
			& \#Opt     & 7 & 6 & 13 & \textbf{14} & 0 & 0 \\
			& \#New     & \textbf{7} & 4 & 6 & 6 & -- & -- \\
			& AVG. Rank & \textbf{2.50} & 3.90 & 3.00 & 2.63 & 5.33 & 3.63  \\ \hline
			
			\multirow{5}{*}{$kC$ snakes (30 inst.)}
			& \#Solved  & \textbf{30} & 16 & 16 & 19 & \textbf{30} & \textbf{30} \\
			& \#Best    & 18 & 7 & 10 & 11 & 2 & \textbf{19}  \\
			& \#Opt     & 7 & 5 & \textbf{10} & \textbf{10} & 0 & 0 \\
			& \#New     & \textbf{11} & 4 & 5 & 6 & -- & -- \\
			& AVG. Rank & \textbf{2.25} & 4.93 & 4.73 & 4.10 & 3.90 & 2.28  \\ \hline
			
			\multirow{5}{*}{Ladder $L_n$ (15 inst.)}
			& \#Solved  & \textbf{15} & \textbf{15} & \textbf{15} & \textbf{15} & \textbf{15} & \textbf{15} \\
			& \#Best    & \textbf{14} & 6 & 9 & 8 & 1 & 3  \\
			& \#Opt     & 7 & 5 & \textbf{8} & \textbf{8} & 0 & 0 \\
			& \#New     & \textbf{12} & 6 & 7 & 10 & -- & -- \\
			& AVG. Rank & \textbf{1.93} & 3.63 & 3.53 & 2.63 & 5.50 & 3.77  \\ \hline
			
			\multirow{5}{*}{Book $B_n$ (15 inst.)}
			& \#Solved  & \textbf{15} & \textbf{15} & \textbf{15} & \textbf{15} & \textbf{15} & \textbf{15} \\
			& \#Best    & \textbf{15} & \textbf{15} & \textbf{15} & \textbf{15} & 6 & 7  \\
			& \#Opt     & 13 & 8 & \textbf{15} & \textbf{15} & 0 & 0 \\
			& \#New     & \textbf{8} & \textbf{8} & \textbf{8} & \textbf{8} & -- & -- \\
			& AVG. Rank & \textbf{2.93} & \textbf{2.93} & \textbf{2.93} & \textbf{2.93} & 4.87 & 4.40  \\ \hline
			
			\multirow{5}{*}{Friend.\ $F_n$ (15 inst.)}
			& \#Solved  & \textbf{15} & \textbf{15} & \textbf{15} & \textbf{15} & \textbf{15} & \textbf{15} \\
			& \#Best    & \textbf{15} & \textbf{15} & \textbf{15} & \textbf{15} & 7 & \textbf{15}  \\
			& \#Opt     & 8 & 7 & \textbf{15} & \textbf{15} & 0 & 0 \\
			& \#New     & 0 & 0 & 0 & 0 & -- & -- \\
			& AVG. Rank & \textbf{3.23} & \textbf{3.23} & \textbf{3.23} & \textbf{3.23} & 4.83 & \textbf{3.23}  \\ \hline
			
			\multirow{5}{*}{Bin.\ tree $BT_k$ (8 inst.)}
			& \#Solved  & \textbf{8} & 6 & 6 & 7 & \textbf{8} & \textbf{8} \\
			& \#Best    & 6 & 5 & 6 & 6 & 6 & \textbf{8}  \\
			& \#Opt     & 3 & 3 & \textbf{6} & \textbf{6} & 0 & 0 \\
			& \#New     & 0 & 0 & 0 & 0 & -- & -- \\
			& AVG. Rank & 3.31 & 4.44 & 4.06 & 3.81 & 3.06 & \textbf{2.81}  \\ \hline
			\hline
			
			\multirow{5}{*}{\textbf{All} (98 inst.)}
			& \#Solved  & \textbf{98} & 82 & 82 & 86 & \textbf{98} & \textbf{98} \\
			& \#Best    & \textbf{83} & 56 & 68 & 69 & 24 & 60  \\
			& \#Opt     & 45 & 34 & 67 & \textbf{68} & 0 & 0 \\
			& \#New     & \textbf{38} & 22 & 26 & 30 & -- & -- \\
			& AVG. Rank & \textbf{2.58} & 3.97 & 3.72 & 3.31 & 4.59 & 3.23  \\ \hline
		\end{tabularx}
		\vspace{2pt}
		{\footnotesize \#Opt is $0$ for $\mathrm{Ub}_{1}$~\cite{saha2012graph} and $\mathrm{Ub}_{3}$~\cite{badr2020upper}, as neither heuristic carries an optimality certificate. \#New is defined relative to $\min(\mathrm{Ub}_1,\mathrm{Ub}_3)$.}
	\end{table}
 
    In the bounded-diameter regime (book and friendship), the ordering reverses. With diameter fixed at $3$ and $2$ respectively, the radio constraint spans only a few separation values, keeping the ILP conflict matrix sparse; ILP CPLEX solves all $30$ instances to certified optimality in seconds (under $14$\,s even on the largest friendship graph). Order-SAT, dominated by a vertex-label variable count that grows linearly with the span, times out on $B_{14}$--$B_{15}$ (encodings exceed $16$ million clauses) and on $F_9$ onward. Where both terminate, they agree exactly, confirming $rn(B_n)=2n+c_n$ with $c_n\in\{2,3,4,5\}$ for $1\le n\le15$. Against the heuristics, Order-SAT improves $\mathrm{Ub}_{3}$ on $8/15$ book instances (ties the rest) and matches it on all friendship instances, while improving $\mathrm{Ub}_{1}$ on $9$ book and $8$ friendship instances. The direct-encoding variant behaves similarly on both bounded-diameter families when solving all $15/15$ book and $15/15$ friendship instances and ties Order-SAT and ILP CPLEX on \#Best (15/15 on both families), but certifies fewer optima than Order-SAT ($8$ vs.\ $13$ on book, $7$ vs.\ $8$ on friendship) because its larger clause count slows down the final UNSAT proof even when the optimal incumbent is already found.

    In the growing-diameter regime (snakes, ladders, binomial trees), the proposed Order-SAT method is consistently the strongest. On the triangular snake $\Delta_k$ it certifies $7/15$ optima (up to $\Delta_7$ in $75.23$\,s), and beyond that its incumbent already matches the value later certified by ILP CPLEX (e.g.\ $rn(\Delta_{12})=145$, $rn(\Delta_{13})=170$), only the proof is missing. It beats $\mathrm{Ub}_{3}$ on the seven odd-index instances and $\mathrm{Ub}_{1}$ on $13/15$. 
    The $kC_4$ and $kC_6$ subfamilies scale with cycle length $m$: fewer instances are certified as $m$ grows ($4/15$ and $3/15$), yet Order-SAT still meets or beats $\mathrm{Ub}_{3}$ on $12/15$ $kC_4$ instances, falling behind the family-specific heuristic of~\cite{badr2020upper} only on the largest $kC_6$ cases ($k\ge10$). The direct-encoding variant is less scalable, on the $kC_4$/$kC_6$ snakes it solves only $16/30$ instances combined (running out of memory for $kC_4$-snakes with $k\ge11$ and $kC_6$-snakes with $k\ge7$, the same thresholds at which ILP CPLEX also exhausts memory), whereas on the triangular snake it solves all $15/15$ instances but certifies only $6/15$ optima against Order-SAT's $7/15$.

    Across the regime, ILP CPLEX's time-limit incumbents are often far looser than Order-SAT's (e.g.\ $327$ vs.\ $251$ on $L_{15}$) and it exhausts memory on the larger $kC_4$/$kC_6$ snakes and on $BT_6$--$BT_7$, while Order-SAT keeps producing feasible labellings throughout.

    The proposed Order-SAT method is thus the method of choice whenever the graph diameter grows with instance size, both certifying more optima and keeping feasible incumbents tight when proofs are out of reach, while the binary ILP baseline (CPLEX or Gurobi) is preferable for bounded-diameter families where its compact constraint matrix is decisive. The two approaches are therefore complementary, and their combination certifies optimality on $109$ of the $146$ benchmark instances.
	
    \subsection{Summary}

    Across all $146$ benchmark instances, Order-SAT returns the tightest span on $130$ instances and certifies exact optimality on $81$, the best figures among the six compared methods. Combined with the ILP baselines, the framework certifies optimality on $109$ instances overall.

    The Friedman test over all six methods, pooled across all $146$ instances, assigns mean ranks of $2.61$ (Order-SAT), $3.81$ (Direct-SAT), $3.56$ (ILP CPLEX), $3.28$ (ILP Gurobi), $4.70$ ($\mathrm{Ub}_{1}$), and $3.31$ ($\mathrm{Ub}_{3}$), with test statistic $\chi^2_F = 181.02$ and $p < 10^{-36}$, confirming that Order-SAT achieves the best overall ranking, ahead of ILP Gurobi, $\mathrm{Ub}_{3}$, ILP CPLEX, and Direct-SAT, and that the differences are statistically significant.

Relative to the state of the art prior to this work, the proposed framework establishes $38$ new best-known radio numbers. Every instance where the ILP baselines alone find an improvement over $\mathrm{Ub}_1$/$\mathrm{Ub}_3$ is also matched or improved by Order-SAT (Section~\ref{subsec:other-families}), while on the $48$ path and cycle instances with a known closed-form optimum, Order-SAT independently recovers the proven value on $46$ of $48$ cases. Taken together, these results position incremental order-encoding SAT as a practical complement to, rather than a replacement for, commercial ILP solvers that decisive on growing-diameter families where ILP's conflict matrix explodes, and a fast independent certifier wherever ILP already succeeds.

	\section{Conclusions}\label{sec:conclude}

This paper presented an exact incremental SAT framework for the radio $k$-labeling problem that combines a compact order encoding with an incremental optimization strategy. By progressively tightening the admissible labeling span while reusing learned clauses across successive SAT calls, the proposed framework avoids rebuilding the SAT model and significantly improves optimization efficiency.

The experimental evaluation demonstrates that the proposed framework substantially advances the state of the art by establishing 38 new best-known radio numbers among the 146 benchmark instances from nine graph families, and, on the path and cycle families whose radio number is already known in closed form, independently recovering the proven-optimal value on 46 of 48 instances. Overall, the best-known radio number is achieved on 84 of the 146 benchmark instances. Furthermore, when combined with ILP baselines solved by CPLEX and Gurobi, the proposed approach certifies optimal solutions for 109 of the 146 benchmark instances, substantially expanding the collection of benchmark instances with proven optimality. Compared with existing heuristic methods, a direct-encoding SAT formulation, and ILP models solved by commercial optimizers, the proposed framework consistently achieves tighter feasible bounds and superior scalability on graph families whose diameter grows with graph size. In contrast, ILP models remains more effective on bounded-diameter graph families, indicating that SAT and ILP provide complementary strengths for exact radio $k$-labeling optimization.

The proposed symmetry breaking constraints for cycle graphs further enhance the efficiency of the SAT framework by reducing the search space and simplifying the SAT encoding, resulting in substantial reductions in solving time on several benchmark instances. These results demonstrate that exploiting graph symmetries can significantly improve the practical performance of SAT-based exact optimization.

Future work will focus on developing symmetry breaking techniques for additional graph families and extending the proposed incremental SAT framework to other graph labeling problems where exact optimization remains computationally challenging. Overall, this work demonstrates that incremental SAT solving provides a practical and highly competitive exact optimization framework for radio $k$-labeling, while highlighting its broader potential for graph labeling and combinatorial optimization problems.

\backmatter

\section*{Declarations}

\subsection*{Funding}
This work has been supported by VNU University of Engineering and Technology under project number CN26.01.

\subsection*{Competing interests}
The authors declare that they have no competing interests.

\subsection*{Availability of data and materials}
The source code and experimental data supporting the findings of this research are openly available on GitHub at: \url{https://github.com/LO2D168/Radio_K_Labeling}.

	\begin{appendices}
		
		\section{Detailed Per-Instance Results}\label{app:detailed}
		
		This appendix reports the full per-instance comparison underlying the family-level summary of Table~\ref{tab:summary}, extended with the two families (paths and cycles) already summarized in Table~\ref{tab:friedman-pathcycle}. For each of the nine graph families, a single table lists the radio number $rn$, solver status (\textbf{OPT} or \textbf{TO}), and running time obtained by the proposed order-encoding SAT method, by each of the three direct-encoding at-most-one variants (pairwise, sequential, cardinality-network), and by the ILP model solved with CPLEX and with Gurobi, alongside the heuristic upper bounds $\mathrm{Ub}_{1}$~\cite{saha2012graph} and $\mathrm{Ub}_{3}$~\cite{badr2020upper}. The sequential variant is the best-performing of the three direct-encoding variants and is the one reported as ``Direct-SAT'' in Table~\ref{tab:summary}.
		
		\begin{landscape}
			\begin{table}[p]
				\centering
				\caption{Experimental results for paths $P_n$.}\label{tab:compare-path-full}
					\tiny
					\setlength{\tabcolsep}{2pt}
					\renewcommand{\arraystretch}{1.25}
					\begin{tabular}{cc rcr rcr rcr rcr rcr rcr r r r}
						\toprule
						\multirow{2}{*}{$n$} & \multirow{2}{*}{$|V|$}
						& \multicolumn{3}{c}{Order-SAT}
						& \multicolumn{3}{c}{Direct-SAT (pairwise)}
						& \multicolumn{3}{c}{Direct-SAT (seqcounter)}
						& \multicolumn{3}{c}{Direct-SAT (cardnetwrk))}
						& \multicolumn{3}{c}{ILP CPLEX}
						& \multicolumn{3}{c}{ILP Gurobi}
						& \multirow{2}{*}{$\mathrm{Ub}_{0}$~\cite{LiuZhu2005}}
						& \multirow{2}{*}{$\mathrm{Ub}_{1}$~\cite{saha2012graph}}
						& \multirow{2}{*}{$\mathrm{Ub}_{3}$~\cite{badr2020upper}} \\
						\cmidrule(lr){3-5}\cmidrule(lr){6-8}\cmidrule(lr){9-11}\cmidrule(lr){12-14}\cmidrule(lr){15-17}\cmidrule(lr){18-20}
						& & $rn$ & St. & Time (s) & $rn$ & St. & Time (s) & $rn$ & St. & Time (s) & $rn$ & St. & Time (s) & $rn$ & St. & Time (s) & $rn$ & St. & Time (s) & & & \\
						\midrule
						1 & 1 & 0 & OPT & 0.49 & 0 & TO & 0.02 & 0 & TO & 0.03 & 0 & TO & 0.02 & 0 & OPT & 0.03 & 0 & OPT & 0.02 & 0 & 0 & 0 \\
						2 & 2 & 1 & OPT & 0.12 & 2 & TO & 1200.00 & 1 & OPT & 0.02 & 1 & OPT & 0.04 & 1 & OPT & 0.20 & 1 & OPT & 0.12 & 1 & 1 & 1 \\
						3 & 3 & 3 & OPT & 0.10 & 5 & TO & 1200.01 & 3 & OPT & 0.03 & 3 & OPT & 0.03 & 3 & OPT & 0.11 & 3 & OPT & 0.03 & 3 & 3 & 3 \\
						4 & 4 & 5 & OPT & 0.16 & 7 & TO & 1200.00 & 5 & OPT & 0.02 & 5 & OPT & 0.03 & 5 & OPT & 0.08 & 5 & OPT & 0.03 & 5 & 6 & 5 \\
						5 & 5 & 10 & OPT & 0.19 & 13 & TO & 1200.00 & 10 & OPT & 0.05 & 10 & OPT & 0.05 & 10 & OPT & 0.24 & 10 & OPT & 0.07 & 10 & 11 & 10 \\
						6 & 6 & 13 & OPT & 0.27 & 16 & TO & 1200.00 & 13 & OPT & 0.08 & 13 & OPT & 0.09 & 13 & OPT & 0.24 & 13 & OPT & 0.10 & 13 & 15 & 14 \\
						7 & 7 & 20 & OPT & 0.28 & 25 & TO & 1200.00 & 20 & OPT & 0.33 & 20 & OPT & 0.35 & 20 & OPT & 0.51 & 20 & OPT & 0.41 & 20 & 22 & 20 \\
						8 & 8 & 25 & OPT & 0.38 & 29 & TO & 1200.01 & 25 & OPT & 1.25 & 25 & OPT & 1.12 & 25 & OPT & 0.94 & 25 & OPT & 0.40 & 25 & 28 & 26 \\
						9 & 9 & 34 & OPT & 0.64 & 41 & TO & 1200.00 & 34 & OPT & 4.30 & 34 & OPT & 4.90 & 34 & OPT & 2.49 & 34 & OPT & 2.19 & 34 & 37 & 35 \\
						10 & 10 & 41 & OPT & 1.24 & 46 & TO & 1200.01 & 41 & OPT & 34.69 & 41 & OPT & 15.93 & 41 & OPT & 3.20 & 41 & OPT & 1.82 & 41 & 45 & 42 \\
						11 & 11 & 52 & OPT & 2.84 & 61 & TO & 1200.01 & 52 & OPT & 211.27 & 52 & OPT & 234.16 & 52 & OPT & 7.90 & 52 & OPT & 7.42 & 52 & 56 & 53 \\
						12 & 12 & 61 & OPT & 4.28 & 67 & TO & 1200.01 & 61 & OPT & 1077.67 & 61 & TO & 1220.95 & 61 & OPT & 8.63 & 61 & OPT & 7.42 & 61 & 66 & 62 \\
						13 & 13 & 74 & OPT & 22.66 & 85 & TO & 1200.42 & 74 & TO & 1200.36 & 74 & TO & 1223.02 & 74 & OPT & 21.20 & 74 & OPT & 33.07 & 74 & 80 & 75 \\
						14 & 14 & 85 & OPT & 67.52 & 92 & TO & 1200.00 & 85 & TO & 1200.25 & 85 & TO & 1200.06 & 85 & OPT & 19.02 & 85 & OPT & 35.06 & 85 & 91 & 86 \\
						15 & 15 & 100 & OPT & 597.51 & 113 & TO & 1200.01 & 101 & TO & 1200.89 & 101 & TO & 1217.10 & 100 & OPT & 80.50 & 100 & OPT & 118.31 & 100 & 107 & 101 \\
						16 & 16 & 113 & TO & 1213.36 & 121 & TO & 1200.03 & 113 & TO & 1200.69 & 115 & TO & 1208.47 & 113 & OPT & 49.93 & 113 & OPT & 49.10 & 113 & 120 & 114 \\
						17 & 17 & 130 & TO & 1200.02 & 145 & TO & 1200.04 & 133 & TO & 1200.89 & 131 & TO & 1209.69 & 130 & OPT & 433.15 & 130 & OPT & 367.89 & 130 & 138 & 132 \\
						18 & 18 & 145 & TO & 1200.11 & 154 & TO & 1200.01 & 146 & TO & 1201.48 & 148 & TO & 1219.49 & 145 & OPT & 408.27 & 145 & OPT & 269.02 & 145 & 153 & 146 \\
						19 & 19 & 164 & TO & 1200.13 & 181 & TO & 1200.03 & 167 & TO & 1200.19 & 169 & TO & 1209.86 & 164 & TO & 1270.57 & 164 & OPT & 578.88 & 164 & 173 & 166 \\
						20 & 20 & 181 & TO & 1200.12 & 191 & TO & 1200.06 & 183 & TO & 1200.26 & 188 & TO & 1221.55 & 181 & OPT & 421.89 & 181 & OPT & 573.47 & 181 & 190 & 182 \\
						21 & 21 & 202 & TO & 1200.11 & 221 & TO & 1200.11 & 208 & TO & 1200.18 & 211 & TO & 1220.57 & 202 & TO & 1306.78 & 203 & TO & 1234.00 & 202 & 213 & 204 \\
						22 & 22 & 221 & TO & 1200.10 & 232 & TO & 1200.07 & 227 & TO & 1200.30 & 226 & TO & 1207.55 & 227 & TO & 1339.84 & 229 & TO & 1236.59 & 221 & 231 & 222 \\
						23 & 23 & 244 & TO & 1200.05 & 265 & TO & 1200.12 & 253 & TO & 1200.35 & 507 & TO & 1200.79 & 254 & TO & 1383.31 & 250 & TO & 1246.00 & 244 & 256 & 246 \\
						24 & 24 & 265 & TO & 1200.51 & 277 & TO & 1200.33 & 276 & TO & 1200.48 & -- & -- & -- & 276 & TO & 1419.71 & 273 & TO & 1259.60 & 265 & 276 & 266 \\
						25 & 25 & 291 & TO & 1200.06 & -- & -- & -- & 300 & TO & 1200.65 & -- & -- & -- & 312 & TO & 1480.06 & 295 & TO & 1275.62 & 290 & 303 & 293 \\
						\bottomrule
					\end{tabular}
			\end{table}
		
			\begin{table}[p]
				\centering
				\caption{Experimental results for cycles $C_n$.}\label{tab:compare-cycle-full}
					\tiny
					\setlength{\tabcolsep}{1.6pt}
					\renewcommand{\arraystretch}{1.25}
					\begin{tabular}{cc rcr rcr rcr rcr rcr rcr rcr r r r}
						\toprule
						\multirow{2}{*}{$n$} & \multirow{2}{*}{$|V|$}
						& \multicolumn{3}{c}{Order-SAT}
						& \multicolumn{3}{c}{Direct-SAT (pairwise)}
						& \multicolumn{3}{c}{Direct-SAT (seqcounter)}
						& \multicolumn{3}{c}{Direct-SAT (cardnetwrk))}
						& \multicolumn{3}{c}{ILP CPLEX}
						& \multicolumn{3}{c}{ILP Gurobi}
						& \multirow{2}{*}{$\mathrm{Ub}_{0}$~\cite{LiuZhu2005}}
						& \multirow{2}{*}{$\mathrm{Ub}_{1}$~\cite{saha2012graph}}
						& \multirow{2}{*}{$\mathrm{Ub}_{3}$~\cite{badr2020upper}} \\
						\cmidrule(lr){3-5}\cmidrule(lr){6-8}\cmidrule(lr){9-11}\cmidrule(lr){12-14}\cmidrule(lr){15-17}\cmidrule(lr){18-20}\cmidrule(lr){21-23}
						& & $rn$ & St. & Time (s) & $rn$ & St. & Time (s) & $rn$ & St. & Time (s) & $rn$ & St. & Time (s) & $rn$ & St. & Time (s) & $rn$ & St. & Time (s) & $rn$ & St. & Time (s) & & & \\
						\midrule
						3 & 3 & 2 & OPT & 0.12 & 2 & OPT & 0.09 & 2 & OPT & 0.04 & 2 & OPT & 3.45 & 2 & OPT & 0.04 & 2 & OPT & 0.06 & 2 & OPT & 0.01 & 2 & 2 & 2 \\
						4 & 4 & 4 & OPT & 0.07 & 4 & OPT & 0.06 & 4 & OPT & 0.04 & 4 & OPT & 0.33 & 4 & OPT & 0.03 & 4 & OPT & 0.05 & 4 & OPT & 0.03 & 4 & 4 & 4 \\
						5 & 5 & 4 & OPT & 0.14 & 4 & OPT & 0.09 & 4 & OPT & 0.03 & 4 & OPT & 0.20 & 4 & OPT & 0.02 & 4 & OPT & 0.06 & 4 & OPT & 0.02 & 4 & 4 & 4 \\
						6 & 6 & 7 & OPT & 0.21 & 7 & OPT & 0.13 & 7 & OPT & 0.04 & 7 & OPT & 0.30 & 7 & OPT & 0.04 & 7 & OPT & 0.14 & 7 & OPT & 0.04 & 7 & 7 & 7 \\
						7 & 7 & 9 & OPT & 0.20 & 9 & OPT & 0.17 & 9 & OPT & 0.05 & 9 & OPT & 0.35 & 9 & OPT & 0.07 & 9 & OPT & 0.29 & 9 & OPT & 0.17 & 9 & 10 & 9 \\
						8 & 8 & 13 & OPT & 0.28 & 13 & OPT & 0.23 & 13 & OPT & 0.11 & 13 & OPT & 0.36 & 13 & OPT & 0.15 & 13 & OPT & 0.70 & 13 & OPT & 0.50 & 13 & 13 & 13 \\
						9 & 9 & 12 & OPT & 0.37 & 12 & OPT & 0.25 & 12 & OPT & 0.11 & 12 & OPT & 0.38 & 12 & OPT & 0.18 & 12 & OPT & 0.50 & 12 & OPT & 0.40 & 12 & 13 & 12 \\
						10 & 10 & 17 & OPT & 0.56 & 17 & OPT & 0.39 & 17 & OPT & 0.32 & 17 & OPT & 1.06 & 17 & OPT & 0.57 & 17 & OPT & 1.85 & 17 & OPT & 1.25 & 17 & 17 & 17 \\
						11 & 11 & 20 & OPT & 1.00 & 20 & OPT & 0.51 & 20 & OPT & 2.55 & 20 & OPT & 4.48 & 20 & OPT & 3.46 & 20 & OPT & 7.32 & 20 & OPT & 7.64 & 20 & 21 & 20 \\
						12 & 12 & 26 & OPT & 2.02 & 26 & OPT & 0.70 & 26 & OPT & 5.69 & 26 & OPT & 9.79 & 26 & OPT & 9.34 & 26 & OPT & 12.60 & 26 & OPT & 21.96 & 26 & 26 & 26 \\
						13 & 13 & 24 & OPT & 1.32 & 24 & OPT & 0.72 & 24 & OPT & 2.25 & 24 & OPT & 4.73 & 24 & OPT & 5.05 & 24 & OPT & 7.60 & 24 & OPT & 9.98 & 24 & 25 & 24 \\
						14 & 14 & 31 & OPT & 2.26 & 31 & OPT & 1.21 & 31 & OPT & 8.34 & 31 & OPT & 16.40 & 31 & OPT & 21.35 & 31 & OPT & 15.11 & 31 & OPT & 49.47 & 31 & 31 & 31 \\
						15 & 15 & 35 & OPT & 44.31 & 35 & OPT & 3.49 & 35 & OPT & 1049.48 & 35 & OPT & 925.44 & 35 & OPT & 1155.01 & 35 & OPT & 612.25 & 35 & OPT & 750.52 & 35 & 36 & 35 \\
						16 & 16 & 43 & OPT & 95.06 & 43 & OPT & 4.75 & 43 & TO & 1200.20 & 43 & TO & 1200.17 & 43 & TO & 1200.16 & 43 & OPT & 1028.00 & 43 & TO & 1201.67 & 43 & 43 & 43 \\
						17 & 17 & 40 & OPT & 7.56 & 40 & OPT & 2.23 & 40 & OPT & 237.17 & 40 & OPT & 165.32 & 40 & OPT & 218.92 & 40 & OPT & 47.55 & 40 & OPT & 245.95 & 40 & 44 & 40 \\
						18 & 18 & 49 & OPT & 34.00 & 49 & OPT & 6.90 & 49 & OPT & 1018.65 & 49 & TO & 1200.18 & 49 & TO & 1200.11 & 49 & OPT & 126.21 & 49 & OPT & 716.23 & 49 & 49 & 49 \\
						19 & 19 & 54 & TO & 1201.95 & 54 & OPT & 183.77 & 54 & TO & 1200.16 & 54 & TO & 1200.17 & 54 & TO & 1201.14 & 54 & TO & 1222.86 & 54 & TO & 1203.69 & 54 & 55 & 54 \\
						20 & 20 & 64 & TO & 1200.60 & 64 & OPT & 298.08 & 64 & TO & 1200.13 & 64 & TO & 1200.18 & 64 & TO & 1200.26 & 64 & TO & 1220.97 & 65 & TO & 1205.49 & 64 & 64 & 64 \\
						21 & 21 & 60 & OPT & 203.95 & 60 & OPT & 11.00 & 60 & TO & 1200.59 & 60 & TO & 1200.19 & 60 & TO & 1204.32 & 60 & OPT & 426.67 & 60 & TO & 1206.38 & 60 & 64 & 60 \\
						22 & 22 & 71 & TO & 1209.23 & 71 & OPT & 188.79 & 71 & TO & 1200.16 & 74 & TO & 1200.20 & 77 & TO & 1206.81 & 71 & TO & 1233.56 & 75 & TO & 1208.45 & 71 & 71 & 71 \\
						23 & 23 & 77 & TO & 1215.97 & 77 & TO & 1217.37 & 77 & TO & 1200.16 & 79 & TO & 1200.20 & 81 & TO & 1205.83 & 80 & TO & 1241.36 & 78 & TO & 1210.30 & 77 & 78 & 77 \\
						24 & 24 & 89 & TO & 1203.63 & 90 & TO & 1207.48 & 93 & TO & 1200.22 & 96 & TO & 1200.20 & 99 & TO & 1202.00 & 114 & TO & 1251.98 & 99 & TO & 1213.66 & 89 & 89 & 89 \\
						25 & 25 & 84 & TO & 1216.02 & 84 & OPT & 870.49 & 85 & TO & 1204.13 & 92 & TO & 1200.28 & 97 & TO & 1207.72 & 93 & TO & 1254.75 & 95 & TO & 1215.06 & 84 & 89 & 84 \\
						\bottomrule
					\end{tabular}
			\end{table}
		
			\begin{table}[p]
				\centering
				\caption{Experimental results for triangular snakes $\Delta_k$.}\label{tab:compare-trisnake-full}
				\tiny
				\setlength{\tabcolsep}{2pt}
				\renewcommand{\arraystretch}{1.25}
				\begin{tabular}{cc rcr rcr rcr rcr rcr rcr r r}
					\toprule
					\multirow{2}{*}{$k$} & \multirow{2}{*}{$|V|$}
					& \multicolumn{3}{c}{Order-SAT}
					& \multicolumn{3}{c}{Direct-SAT (pairwise)}
					& \multicolumn{3}{c}{Direct-SAT (seqcounter)}
					& \multicolumn{3}{c}{Direct-SAT (cardnetwrk)}
					& \multicolumn{3}{c}{ILP CPLEX}
					& \multicolumn{3}{c}{ILP Gurobi}
					& \multirow{2}{*}{$\mathrm{Ub}_{1}$~\cite{saha2012graph}}
					& \multirow{2}{*}{$\mathrm{Ub}_{3}$~\cite{badr2020upper}} \\
					\cmidrule(lr){3-5}\cmidrule(lr){6-8}\cmidrule(lr){9-11}\cmidrule(lr){12-14}\cmidrule(lr){15-17}\cmidrule(lr){18-20}
					& & $rn$ & St. & Time (s) & $rn$ & St. & Time (s) & $rn$ & St. & Time (s) & $rn$ & St. & Time (s) & $rn$ & St. & Time (s) & $rn$ & St. & Time (s) & & \\
					\midrule
					1 & 3 & 2 & OPT & 0.07 & 2 & OPT & 0.54 & 2 & OPT & 0.65 & 2 & OPT & 0.87 & 2 & OPT & 0.04 & 2 & OPT & 0.02 & 2 & 2 \\
					2 & 5 & 5 & OPT & 0.08 & 5 & OPT & 0.14 & 5 & OPT & 0.20 & 5 & OPT & 0.13 & 5 & OPT & 0.06 & 5 & OPT & 0.02 & 5 & 5 \\
					3 & 7 & 10 & OPT & 0.14 & 10 & OPT & 0.18 & 10 & OPT & 0.26 & 10 & OPT & 0.19 & 10 & OPT & 0.16 & 10 & OPT & 0.07 & 11 & 11 \\
					4 & 9 & 17 & OPT & 0.25 & 17 & OPT & 0.34 & 17 & OPT & 0.60 & 17 & OPT & 0.55 & 17 & OPT & 0.46 & 17 & OPT & 0.35 & 18 & 17 \\
					5 & 11 & 26 & OPT & 0.69 & 26 & OPT & 3.81 & 26 & OPT & 7.06 & 26 & OPT & 8.56 & 26 & OPT & 4.37 & 26 & OPT & 1.11 & 28 & 28 \\
					6 & 13 & 37 & OPT & 5.71 & 37 & OPT & 227.07 & 37 & OPT & 323.53 & 37 & OPT & 363.80 & 37 & OPT & 4.55 & 37 & OPT & 3.05 & 39 & 37 \\
					7 & 15 & 50 & OPT & 75.23 & 50 & TO & 1219.95 & 50 & TO & 1200.16 & 50 & TO & 1223.71 & 50 & OPT & 8.28 & 50 & OPT & 7.66 & 53 & 52 \\
					8 & 17 & 65 & TO & 1223.44 & 65 & TO & 1221.34 & 65 & TO & 1200.17 & 65 & TO & 1222.79 & 65 & OPT & 17.70 & 65 & OPT & 22.81 & 68 & 65 \\
					9 & 19 & 82 & TO & 2638.93 & 82 & TO & 1220.87 & 83 & TO & 1200.11 & 83 & TO & 1217.70 & 82 & OPT & 42.09 & 82 & OPT & 41.20 & 86 & 85 \\
					10 & 21 & 101 & TO & 1219.28 & 102 & TO & 1215.94 & 102 & TO & 1200.19 & 103 & TO & 1219.90 & 101 & OPT & 99.47 & 101 & OPT & 67.47 & 105 & 101 \\
					11 & 23 & 122 & TO & 1222.68 & 122 & TO & 1205.92 & 125 & TO & 1200.22 & 126 & TO & 1208.91 & 122 & OPT & 153.07 & 122 & OPT & 217.11 & 127 & 125 \\
					12 & 25 & 145 & TO & 1223.69 & 147 & TO & 1218.43 & 149 & TO & 1200.26 & 149 & TO & 1200.32 & 145 & OPT & 830.68 & 145 & OPT & 164.36 & 150 & 145 \\
					13 & 27 & 170 & TO & 1216.62 & 174 & TO & 1200.94 & 175 & TO & 1200.26 & 179 & TO & 1200.48 & 170 & OPT & 479.21 & 170 & OPT & 744.05 & 176 & 174 \\
					14 & 29 & 197 & TO & 1211.06 & 201 & TO & 1205.54 & 206 & TO & 1200.33 & 206 & TO & 1206.73 & 205 & TO & 1344.57 & 197 & OPT & 492.24 & 203 & 197 \\
					15 & 31 & 226 & TO & 1202.15 & 235 & TO & 1211.30 & 238 & TO & 1200.53 & -- & -- & -- & 238 & TO & 1405.20 & 232 & TO & 1258.84 & 233 & 230 \\
					\bottomrule
				\end{tabular}
			\end{table}
		
			\begin{table}[p]
				\centering
				\caption{Experimental results for $kC_4$-snakes.}\label{tab:compare-c4snake-full}
				\tiny
				\setlength{\tabcolsep}{2pt}
				\renewcommand{\arraystretch}{1.25}
				\begin{tabular}{cc rcr rcr rcr rcr rcr rcr r r}
					\toprule
					\multirow{2}{*}{$k$} & \multirow{2}{*}{$|V|$}
					& \multicolumn{3}{c}{Order-SAT}
					& \multicolumn{3}{c}{Direct-SAT (pairwise)}
					& \multicolumn{3}{c}{Direct-SAT (seqcounter)}
					& \multicolumn{3}{c}{Direct-SAT (cardnetwrk)}
					& \multicolumn{3}{c}{ILP CPLEX}
					& \multicolumn{3}{c}{ILP Gurobi}
					& \multirow{2}{*}{$\mathrm{Ub}_{1}$~\cite{saha2012graph}}
					& \multirow{2}{*}{$\mathrm{Ub}_{3}$~\cite{badr2020upper}} \\
					\cmidrule(lr){3-5}\cmidrule(lr){6-8}\cmidrule(lr){9-11}\cmidrule(lr){12-14}\cmidrule(lr){15-17}\cmidrule(lr){18-20}
					& & $rn$ & St. & Time (s) & $rn$ & St. & Time (s) & $rn$ & St. & Time (s) & $rn$ & St. & Time (s) & $rn$ & St. & Time (s) & $rn$ & St. & Time (s) & & \\
					\midrule
					1 & 4 & 4 & OPT & 0.08 & 4 & OPT & 3.41 & 4 & OPT & 2.82 & 4 & OPT & 0.06 & 4 & OPT & 0.04 & 4 & OPT & 0.02 & 4 & 4 \\
					2 & 7 & 15 & OPT & 0.15 & 15 & OPT & 0.51 & 15 & OPT & 0.37 & 15 & OPT & 0.21 & 15 & OPT & 0.25 & 15 & OPT & 0.13 & 17 & 15 \\
					3 & 10 & 31 & OPT & 0.56 & 31 & OPT & 6.35 & 31 & OPT & 13.84 & 31 & OPT & 18.04 & 31 & OPT & 2.66 & 31 & OPT & 2.66 & 33 & 31 \\
					4 & 13 & 53 & OPT & 12.53 & 53 & TO & 1207.42 & 53 & TO & 1200.16 & 53 & TO & 1244.45 & 53 & OPT & 6.50 & 53 & OPT & 10.95 & 59 & 55 \\
					5 & 16 & 81 & TO & 1216.80 & 82 & TO & 1200.21 & 82 & TO & 1200.23 & 82 & TO & 1239.81 & 81 & OPT & 68.14 & 81 & OPT & 173.20 & 86 & 83 \\
					6 & 19 & 115 & TO & 1203.07 & 116 & TO & 1200.30 & 118 & TO & 1200.22 & 118 & TO & 1225.73 & 115 & OPT & 145.40 & 115 & OPT & 111.43 & 124 & 117 \\
					7 & 22 & 155 & TO & 1200.11 & 158 & TO & 1200.36 & 162 & TO & 1200.27 & 160 & TO & 1204.44 & 158 & TO & 1261.54 & 155 & TO & 1217.90 & 163 & 157 \\
					8 & 25 & 202 & TO & 1200.10 & 207 & TO & 1200.42 & 211 & TO & 1200.40 & 211 & TO & 1204.60 & 208 & TO & 1325.68 & 205 & TO & 1233.14 & 214 & 205 \\
					9 & 28 & 254 & TO & 1200.10 & -- & -- & -- & 271 & TO & 1200.56 & -- & -- & -- & 265 & TO & 1431.95 & 259 & TO & 1268.18 & 264 & 257 \\
					10 & 31 & 313 & TO & 1200.10 & -- & -- & -- & 333 & TO & 1204.02 & -- & -- & -- & 352 & TO & 1579.22 & 340 & TO & 1307.11 & 327 & 316 \\
					11 & 34 & 379 & TO & 1200.04 & -- & -- & -- & -- & -- & -- & -- & -- & -- & -- & -- & -- & 413 & TO & 1368.99 & 389 & 379 \\
					12 & 37 & 451 & TO & 1200.13 & -- & -- & -- & -- & -- & -- & -- & -- & -- & -- & -- & -- & 495 & TO & 1454.13 & 465 & 451 \\
					13 & 40 & 531 & TO & 1200.16 & -- & -- & -- & -- & -- & -- & -- & -- & -- & -- & -- & -- & -- & -- & -- & 538 & 527 \\
					14 & 43 & 616 & TO & 1200.18 & -- & -- & -- & -- & -- & -- & -- & -- & -- & -- & -- & -- & -- & -- & -- & 626 & 609 \\
					15 & 46 & 711 & TO & 1200.15 & -- & -- & -- & -- & -- & -- & -- & -- & -- & -- & -- & -- & -- & -- & -- & 711 & 697 \\
					\bottomrule
				\end{tabular}
			\end{table}
		
			\begin{table}[p]
				\centering
				\caption{Experimental results for $kC_6$-snakes.}\label{tab:compare-c6snake-full}
				\tiny
				\setlength{\tabcolsep}{2pt}
				\renewcommand{\arraystretch}{1.25}
				\begin{tabular}{cc rcr rcr rcr rcr rcr rcr r r}
					\toprule
					\multirow{2}{*}{$k$} & \multirow{2}{*}{$|V|$}
					& \multicolumn{3}{c}{Order-SAT}
					& \multicolumn{3}{c}{Direct-SAT (pairwise)}
					& \multicolumn{3}{c}{Direct-SAT (seqcounter)}
					& \multicolumn{3}{c}{Direct-SAT (cardnetwrk)}
					& \multicolumn{3}{c}{ILP CPLEX}
					& \multicolumn{3}{c}{ILP Gurobi}
					& \multirow{2}{*}{$\mathrm{Ub}_{1}$~\cite{saha2012graph}}
					& \multirow{2}{*}{$\mathrm{Ub}_{3}$~\cite{badr2020upper}} \\
					\cmidrule(lr){3-5}\cmidrule(lr){6-8}\cmidrule(lr){9-11}\cmidrule(lr){12-14}\cmidrule(lr){15-17}\cmidrule(lr){18-20}
					& & $rn$ & St. & Time (s) & $rn$ & St. & Time (s) & $rn$ & St. & Time (s) & $rn$ & St. & Time (s) & $rn$ & St. & Time (s) & $rn$ & St. & Time (s) & & \\
					\midrule
					1 & 6 & 7 & OPT & 0.34 & 7 & OPT & 0.06 & 7 & OPT & 0.07 & 7 & OPT & 0.08 & 7 & OPT & 0.28 & 7 & OPT & 0.96 & 7 & 7 \\
					2 & 11 & 35 & OPT & 2.20 & 35 & OPT & 118.28 & 35 & OPT & 197.50 & 35 & OPT & 168.50 & 35 & OPT & 1.78 & 35 & OPT & 1.89 & 38 & 35 \\
					3 & 16 & 71 & OPT & 497.99 & 72 & TO & 1200.06 & 71 & TO & 1200.10 & 72 & TO & 1217.86 & 71 & OPT & 46.74 & 71 & OPT & 80.41 & 77 & 73 \\
					4 & 21 & 129 & TO & 1200.10 & 131 & TO & 1200.18 & 132 & TO & 1200.16 & 133 & TO & 1220.52 & 129 & OPT & 413.07 & 129 & OPT & 269.64 & 139 & 135 \\
					5 & 26 & 196 & TO & 1200.10 & 203 & TO & 1200.33 & 205 & TO & 1200.31 & 206 & TO & 1223.17 & 225 & TO & 1321.92 & 202 & TO & 1233.32 & 204 & 197 \\
					6 & 31 & 284 & TO & 1200.10 & -- & -- & -- & 298 & TO & 1200.67 & -- & -- & -- & 328 & TO & 1496.47 & 289 & TO & 1288.47 & 297 & 286 \\
					7 & 36 & 383 & TO & 1201.52 & -- & -- & -- & -- & -- & -- & -- & -- & -- & -- & -- & -- & 408 & TO & 1397.11 & 391 & 381 \\
					8 & 41 & 508 & TO & 1201.07 & -- & -- & -- & -- & -- & -- & -- & -- & -- & -- & -- & -- & -- & -- & -- & 519 & 507 \\
					9 & 46 & 638 & TO & 1201.72 & -- & -- & -- & -- & -- & -- & -- & -- & -- & -- & -- & -- & -- & -- & -- & 638 & 625 \\
					10 & 51 & 801 & TO & 1201.12 & -- & -- & -- & -- & -- & -- & -- & -- & -- & -- & -- & -- & -- & -- & -- & 795 & 779 \\
					11 & 56 & 955 & TO & 1201.41 & -- & -- & -- & -- & -- & -- & -- & -- & -- & -- & -- & -- & -- & -- & -- & 945 & 929 \\
					12 & 61 & 1156 & TO & 1202.71 & -- & -- & -- & -- & -- & -- & -- & -- & -- & -- & -- & -- & -- & -- & -- & 1136 & 1118 \\
					13 & 66 & 1331 & TO & 1208.20 & -- & -- & -- & -- & -- & -- & -- & -- & -- & -- & -- & -- & -- & -- & -- & 1312 & 1293 \\
					14 & 71 & 1562 & TO & 1200.65 & -- & -- & -- & -- & -- & -- & -- & -- & -- & -- & -- & -- & -- & -- & -- & 1534 & 1509 \\
					15 & 76 & 1766 & TO & 1203.11 & -- & -- & -- & -- & -- & -- & -- & -- & -- & -- & -- & -- & -- & -- & -- & 1739 & 1717 \\
					\bottomrule
				\end{tabular}
			\end{table}
		
			\begin{table}[p]
				\centering
				\caption{Experimental results for ladder graphs $L_n$.}\label{tab:compare-ladder-full}
				\tiny
				\setlength{\tabcolsep}{2pt}
				\renewcommand{\arraystretch}{1.25}
				\begin{tabular}{cc rcr rcr rcr rcr rcr rcr r r}
					\toprule
					\multirow{2}{*}{$k$} & \multirow{2}{*}{$|V|$}
					& \multicolumn{3}{c}{Order-SAT}
					& \multicolumn{3}{c}{Direct-SAT (pairwise)}
					& \multicolumn{3}{c}{Direct-SAT (seqcounter)}
					& \multicolumn{3}{c}{Direct-SAT (cardnetwrk)}
					& \multicolumn{3}{c}{ILP CPLEX}
					& \multicolumn{3}{c}{ILP Gurobi}
					& \multirow{2}{*}{$\mathrm{Ub}_{1}$~\cite{saha2012graph}}
					& \multirow{2}{*}{$\mathrm{Ub}_{3}$~\cite{badr2020upper}}
                    \\
					\cmidrule(lr){3-5}\cmidrule(lr){6-8}\cmidrule(lr){9-11}\cmidrule(lr){12-14}\cmidrule(lr){15-17}\cmidrule(lr){18-20}
					& & $rn$ & St. & Time (s) & $rn$ & St. & Time (s) & $rn$ & St. & Time (s) & $rn$ & St. & Time (s) & $rn$ & St. & Time (s) & $rn$ & St. & Time (s) & & \\
					\midrule
					1 & 4 & 4 & OPT & 1.77 & 4 & OPT & 0.06 & 4 & OPT & 0.05 & 4 & OPT & 0.06 & 4 & OPT & 1.32 & 4 & OPT & 0.71 & 4 & 4 \\
					2 & 6 & 8 & OPT & 0.84 & 8 & OPT & 0.03 & 8 & OPT & 0.03 & 8 & OPT & 0.04 & 8 & OPT & 0.17 & 8 & OPT & 0.04 & 9 & 8 \\
					3 & 8 & 15 & OPT & 1.36 & 15 & OPT & 0.11 & 15 & OPT & 0.14 & 15 & OPT & 0.23 & 15 & OPT & 0.75 & 15 & OPT & 0.43 & 18 & 15 \\
					4 & 10 & 23 & OPT & 2.29 & 23 & OPT & 0.79 & 23 & OPT & 1.36 & 23 & OPT & 1.65 & 23 & OPT & 3.51 & 23 & OPT & 1.76 & 27 & 24 \\
					5 & 12 & 34 & OPT & 4.55 & 34 & OPT & 29.32 & 34 & OPT & 68.50 & 34 & OPT & 88.75 & 34 & OPT & 18.52 & 34 & OPT & 17.77 & 40 & 36 \\
					6 & 14 & 45 & OPT & 14.42 & 45 & TO & 1200.02 & 45 & TO & 1200.11 & 45 & TO & 1200.13 & 45 & OPT & 18.95 & 45 & OPT & 48.51 & 50 & 47 \\
					7 & 16 & 60 & OPT & 421.85 & 61 & TO & 1200.12 & 61 & TO & 1200.02 & 61 & TO & 1200.14 & 60 & OPT & 115.45 & 60 & OPT & 381.60 & 70 & 64 \\
					8 & 18 & 75 & TO & 1215.65 & 76 & TO & 1200.12 & 76 & TO & 1200.13 & 77 & TO & 1200.18 & 75 & OPT & 364.33 & 75 & OPT & 733.91 & 85 & 79 \\
					9 & 20 & 95 & TO & 1487.88 & 96 & TO & 1200.14 & 98 & TO & 1200.04 & 99 & TO & 1200.18 & 96 & TO & 1223.95 & 98 & TO & 1205.92 & 108 & 100 \\
					10 & 22 & 114 & TO & 1200.15 & 116 & TO & 1200.17 & 118 & TO & 1200.17 & 117 & TO & 1200.29 & 113 & TO & 1239.92 & 116 & TO & 1210.87 & 124 & 117 \\
					11 & 24 & 139 & TO & 1355.84 & 143 & TO & 1200.29 & 145 & TO & 1200.19 & 145 & TO & 1200.42 & 149 & TO & 1264.13 & 140 & TO & 1216.64 & 154 & 144 \\
					12 & 26 & 161 & TO & 1200.05 & 168 & TO & 1200.27 & 168 & TO & 1200.18 & 172 & TO & 1201.01 & 176 & TO & 1295.62 & 165 & TO & 1222.93 & 175 & 167 \\
					13 & 28 & 190 & TO & 1200.18 & 198 & TO & 1201.09 & 202 & TO & 1200.32 & 205 & TO & 1200.55 & 216 & TO & 1332.20 & 198 & TO & 1234.58 & 208 & 197 \\
					14 & 30 & 217 & TO & 1200.08 & 226 & TO & 1208.89 & 229 & TO & 1200.35 & 236 & TO & 1203.32 & 278 & TO & 1382.71 & 228 & TO & 1252.90 & 230 & 223 \\
					15 & 32 & 251 & TO & 1200.16 & -- & -- & -- & 275 & TO & 1200.54 & -- & -- & -- & 327 & TO & 1473.68 & 256 & TO & 1271.70 & 270 & 257 \\
					\bottomrule
				\end{tabular}
			\end{table}
		
			\begin{table}[p]
				\centering
				\caption{Experimental results for book graphs $B_n$.}\label{tab:compare-book-full}
				\tiny
				\setlength{\tabcolsep}{2pt}
				\renewcommand{\arraystretch}{1.25}
				\begin{tabular}{cc rcr rcr rcr rcr rcr rcr r r}
					\toprule
					\multirow{2}{*}{$k$} & \multirow{2}{*}{$|V|$}
					& \multicolumn{3}{c}{Order-SAT}
					& \multicolumn{3}{c}{Direct-SAT (pairwise)}
					& \multicolumn{3}{c}{Direct-SAT (seqcounter)}
					& \multicolumn{3}{c}{Direct-SAT (cardnetwrk)}
					& \multicolumn{3}{c}{ILP CPLEX}
					& \multicolumn{3}{c}{ILP Gurobi}
					& \multirow{2}{*}{$\mathrm{Ub}_{1}$~\cite{saha2012graph}}
					& \multirow{2}{*}{$\mathrm{Ub}_{3}$~\cite{badr2020upper}} \\
					\cmidrule(lr){3-5}\cmidrule(lr){6-8}\cmidrule(lr){9-11}\cmidrule(lr){12-14}\cmidrule(lr){15-17}\cmidrule(lr){18-20}
					& & $rn$ & St. & Time (s) & $rn$ & St. & Time (s) & $rn$ & St. & Time (s) & $rn$ & St. & Time (s) & $rn$ & St. & Time (s) & $rn$ & St. & Time (s) & & \\
					\midrule
					1 & 4 & 4 & OPT & 0.30 & 4 & OPT & 0.06 & 4 & OPT & 3.02 & 4 & OPT & 0.07 & 4 & OPT & 0.07 & 4 & OPT & 0.03 & 14 & 4 \\
					2 & 6 & 8 & OPT & 0.50 & 8 & OPT & 0.04 & 8 & OPT & 0.35 & 8 & OPT & 0.06 & 8 & OPT & 0.13 & 8 & OPT & 0.05 & 8 & 8 \\
					3 & 8 & 9 & OPT & 0.73 & 9 & OPT & 0.04 & 9 & OPT & 0.29 & 9 & OPT & 0.07 & 9 & OPT & 0.22 & 9 & OPT & 0.08 & 9 & 9 \\
					4 & 10 & 11 & OPT & 0.91 & 11 & OPT & 0.08 & 11 & OPT & 0.44 & 11 & OPT & 0.20 & 11 & OPT & 0.46 & 11 & OPT & 0.16 & 13 & 12 \\
					5 & 12 & 13 & OPT & 1.17 & 13 & OPT & 0.21 & 13 & OPT & 0.59 & 13 & OPT & 0.44 & 13 & OPT & 0.67 & 13 & OPT & 0.24 & 14 & 14 \\
					6 & 14 & 15 & OPT & 1.48 & 15 & OPT & 0.75 & 15 & OPT & 1.50 & 15 & OPT & 1.41 & 15 & OPT & 0.93 & 15 & OPT & 0.30 & 15 & 15 \\
					7 & 16 & 17 & OPT & 1.88 & 17 & OPT & 6.32 & 17 & OPT & 8.25 & 17 & OPT & 7.67 & 17 & OPT & 1.48 & 17 & OPT & 0.67 & 19 & 18 \\
					8 & 18 & 19 & OPT & 2.61 & 19 & OPT & 85.48 & 19 & OPT & 40.94 & 19 & OPT & 59.25 & 19 & OPT & 1.96 & 19 & OPT & 0.73 & 20 & 20 \\
					9 & 20 & 21 & OPT & 5.28 & 21 & OPT & 847.34 & 21 & TO & 1200.10 & 21 & OPT & 1199.43 & 21 & OPT & 3.15 & 21 & OPT & 0.93 & 21 & 21 \\
					10 & 22 & 23 & OPT & 11.01 & 23 & TO & 1232.70 & 23 & TO & 1200.24 & 23 & TO & 1234.33 & 23 & OPT & 3.70 & 23 & OPT & 1.13 & 25 & 24 \\
					11 & 24 & 25 & OPT & 39.22 & 25 & TO & 1232.74 & 25 & TO & 1200.24 & 25 & TO & 1232.50 & 25 & OPT & 4.31 & 25 & OPT & 1.52 & 26 & 26 \\
					12 & 26 & 27 & OPT & 86.19 & 27 & TO & 1229.06 & 27 & TO & 1200.23 & 27 & TO & 1233.59 & 27 & OPT & 5.84 & 27 & OPT & 1.52 & 27 & 27 \\
					13 & 28 & 29 & OPT & 624.32 & 29 & TO & 1223.99 & 29 & TO & 1200.15 & 29 & TO & 1232.92 & 29 & OPT & 6.29 & 29 & OPT & 2.84 & 31 & 30 \\
					14 & 30 & 31 & TO & 1207.39 & 31 & TO & 1216.58 & 31 & TO & 1200.19 & 31 & TO & 1239.35 & 31 & OPT & 9.55 & 31 & OPT & 3.34 & 32 & 32 \\
					15 & 32 & 33 & TO & 1208.13 & 33 & TO & 1200.14 & 33 & TO & 1200.19 & 33 & TO & 1240.96 & 33 & OPT & 11.08 & 33 & OPT & 4.11 & 33 & 33 \\
					\bottomrule
				\end{tabular}
			\end{table}
		
			\begin{table}[p]
				\centering
				\caption{Experimental results for friendship graphs $F_n$.}\label{tab:compare-friendship-full}
				\tiny
				\setlength{\tabcolsep}{2pt}
				\renewcommand{\arraystretch}{1.25}
				\begin{tabular}{cc rcr rcr rcr rcr rcr rcr r r}
					\toprule
					\multirow{2}{*}{$k$} & \multirow{2}{*}{$|V|$}
					& \multicolumn{3}{c}{Order-SAT}
					& \multicolumn{3}{c}{Direct-SAT (pairwise)}
					& \multicolumn{3}{c}{Direct-SAT (seqcounter)}
					& \multicolumn{3}{c}{Direct-SAT (cardnetwrk)}
					& \multicolumn{3}{c}{ILP CPLEX}
					& \multicolumn{3}{c}{ILP Gurobi}
					& \multirow{2}{*}{$\mathrm{Ub}_{1}$~\cite{saha2012graph}}
					& \multirow{2}{*}{$\mathrm{Ub}_{3}$~\cite{badr2020upper}}
                    \\
					\cmidrule(lr){3-5}\cmidrule(lr){6-8}\cmidrule(lr){9-11}\cmidrule(lr){12-14}\cmidrule(lr){15-17}\cmidrule(lr){18-20}
					& & $rn$ & St. & Time (s) & $rn$ & St. & Time (s) & $rn$ & St. & Time (s) & $rn$ & St. & Time (s) & $rn$ & St. & Time (s) & $rn$ & St. & Time (s) & & \\
					\midrule
					1 & 3 & 2 & OPT & 0.06 & 2 & OPT & 0.15 & 2 & OPT & 0.22 & 2 & OPT & 0.18 & 2 & OPT & 0.05 & 2 & OPT & 0.01 & 4 & 2 \\
					2 & 5 & 5 & OPT & 0.07 & 5 & OPT & 0.06 & 5 & OPT & 0.14 & 5 & OPT & 0.03 & 5 & OPT & 0.09 & 5 & OPT & 0.02 & 5 & 5 \\
					3 & 7 & 7 & OPT & 0.13 & 7 & OPT & 0.10 & 7 & OPT & 0.23 & 7 & OPT & 0.06 & 7 & OPT & 0.17 & 7 & OPT & 0.02 & 8 & 7 \\
					4 & 9 & 9 & OPT & 0.18 & 9 & OPT & 0.15 & 9 & OPT & 0.27 & 9 & OPT & 0.17 & 9 & OPT & 0.29 & 9 & OPT & 0.05 & 9 & 9 \\
					5 & 11 & 11 & OPT & 0.73 & 11 & OPT & 9.81 & 11 & OPT & 7.14 & 11 & OPT & 2.09 & 11 & OPT & 0.27 & 11 & OPT & 0.10 & 12 & 11 \\
					6 & 13 & 13 & OPT & 4.25 & 13 & TO & 1200.10 & 13 & OPT & 28.83 & 13 & OPT & 21.74 & 13 & OPT & 0.34 & 13 & OPT & 0.16 & 13 & 13 \\
					7 & 15 & 15 & OPT & 48.32 & 15 & TO & 1200.12 & 15 & OPT & 668.08 & 15 & TO & 1242.65 & 15 & OPT & 0.45 & 15 & OPT & 0.14 & 16 & 15 \\
					8 & 17 & 17 & OPT & 826.06 & 17 & TO & 1200.12 & 17 & TO & 1200.23 & 17 & TO & 1242.78 & 17 & OPT & 0.65 & 17 & OPT & 0.29 & 17 & 17 \\
					9 & 19 & 19 & TO & 1216.20 & 19 & TO & 1200.14 & 19 & TO & 1200.28 & 19 & TO & 1241.28 & 19 & OPT & 0.84 & 19 & OPT & 0.32 & 20 & 19 \\
					10 & 21 & 21 & TO & 1216.69 & 21 & TO & 1200.10 & 21 & TO & 1200.17 & 21 & TO & 1235.24 & 21 & OPT & 1.08 & 21 & OPT & 0.42 & 21 & 21 \\
					11 & 23 & 23 & TO & 1220.76 & 23 & TO & 1200.02 & 23 & TO & 1200.21 & 23 & TO & 1228.60 & 23 & OPT & 1.33 & 23 & OPT & 0.50 & 24 & 23 \\
					12 & 25 & 25 & TO & 1216.52 & 25 & TO & 1200.11 & 25 & TO & 1200.16 & 25 & TO & 1221.51 & 25 & OPT & 1.66 & 25 & OPT & 0.65 & 25 & 25 \\
					13 & 27 & 27 & TO & 1487.93 & 27 & TO & 1200.12 & 27 & TO & 1200.23 & 27 & TO & 1209.60 & 27 & OPT & 2.44 & 27 & OPT & 0.83 & 28 & 27 \\
					14 & 29 & 29 & TO & 1200.02 & 29 & TO & 1200.10 & 29 & TO & 1200.06 & 29 & TO & 1200.12 & 29 & OPT & 13.70 & 29 & OPT & 1.09 & 29 & 29 \\
					15 & 31 & 31 & TO & 1355.90 & 31 & TO & 1200.59 & 31 & TO & 1203.31 & 31 & TO & 1200.14 & 31 & OPT & 3.62 & 31 & OPT & 1.22 & 32 & 31 \\
					\bottomrule
				\end{tabular}
			\end{table}
		
			\begin{table}[p]
				\centering
				\caption{Experimental results on binomial trees $BT_k$.}\label{tab:compare-bitree-full}
				\tiny
				\setlength{\tabcolsep}{2pt}
				\renewcommand{\arraystretch}{1.25}
				\begin{tabular}{cc rcr rcr rcr rcr rcr rcr r r}
					\toprule
					\multirow{2}{*}{$k$} & \multirow{2}{*}{$|V|$}
					& \multicolumn{3}{c}{Order-SAT}
					& \multicolumn{3}{c}{Direct-SAT (pairwise)}
					& \multicolumn{3}{c}{Direct-SAT (seqcounter)}
					& \multicolumn{3}{c}{Direct-SAT (cardnetwrk)}
					& \multicolumn{3}{c}{ILP CPLEX}
					& \multicolumn{3}{c}{ILP Gurobi}
					& \multirow{2}{*}{$\mathrm{Ub}_{1}$~\cite{saha2012graph}}
					& \multirow{2}{*}{$\mathrm{Ub}_{3}$~\cite{badr2020upper}}
                    \\
					\cmidrule(lr){3-5}\cmidrule(lr){6-8}\cmidrule(lr){9-11}\cmidrule(lr){12-14}\cmidrule(lr){15-17}\cmidrule(lr){18-20}
					& & $rn$ & St. & Time (s) & $rn$ & St. & Time (s) & $rn$ & St. & Time (s) & $rn$ & St. & Time (s) & $rn$ & St. & Time (s) & $rn$ & St. & Time (s) & & \\
					\midrule
					0 & 1 & 0 & TO & 0.03 & 0 & TO & 0.03 & 0 & TO & 0.02 & 0 & TO & 0.03 & 0 & OPT & 0.01 & 0 & OPT & 0.01 & 0 & 0 \\
					1 & 2 & 1 & OPT & 0.05 & 1 & OPT & 0.03 & 1 & OPT & 0.02 & 1 & OPT & 0.04 & 1 & OPT & 0.02 & 1 & OPT & 0.01 & 1 & 1 \\
					2 & 4 & 5 & OPT & 0.09 & 5 & OPT & 0.04 & 5 & OPT & 0.02 & 5 & OPT & 0.04 & 5 & OPT & 0.04 & 5 & OPT & 0.02 & 5 & 5 \\
					3 & 8 & 19 & OPT & 0.43 & 19 & OPT & 0.26 & 19 & OPT & 0.37 & 19 & OPT & 0.57 & 19 & OPT & 0.47 & 19 & OPT & 0.16 & 19 & 19 \\
					4 & 16 & 57 & TO & 1200.14 & 57 & TO & 1200.58 & 57 & TO & 1200.64 & 57 & TO & 1200.13 & 57 & OPT & 8.87 & 57 & OPT & 3.16 & 57 & 57 \\
					5 & 32 & 151 & TO & 1200.12 & 156 & TO & 1200.12 & 157 & TO & 1200.29 & -- & -- & -- & 151 & OPT & 229.35 & 151 & OPT & 133.96 & 151 & 151 \\
					6 & 64 & 382 & TO & 1200.25 & -- & -- & -- & -- & -- & -- & -- & -- & -- & -- & -- & -- & 395 & TO & 1526.81 & 375 & 373 \\
					7 & 128 & 1542 & TO & 1200.15 & -- & -- & -- & -- & -- & -- & -- & -- & -- & -- & -- & -- & -- & -- & -- & 887 & 883 \\
					\bottomrule
				\end{tabular}
			\end{table}
		\end{landscape}

		\clearpage
	\end{appendices}
	
	\backmatter
	
	\bibliography{sn-bibliography}

@manual{cplex,
  title = {IBM ILOG CPLEX Optimization Studio User's Manual},
  author = {IBM},
  organization = {IBM Corporation},
  year = {2023},
  url = {https://www.ibm.com/docs/en/icos/22.1.1}
}

@article{badr2020upper,
  title={An upper bound of radio k-coloring problem and its integer linear programming model},
  author={Badr, Elsayed M and Moussa, Mahmoud I},
  journal={Wireless Networks},
  volume={26},
  number={7},
  pages={4955--4964},
  year={2020},
  publisher={Springer}
}

@inproceedings{saha2012graph,
  title={A graph radio k-coloring algorithm},
  author={Saha, Laxman and Panigrahi, Pratima},
  booktitle={International Workshop on Combinatorial Algorithms},
  pages={125--129},
  year={2012},
  organization={Springer}
}

@book{biere2009handbook,
  title={Handbook of satisfiability},
  author={Biere, Armin and van Maaren, Hans and Walsh, Toby},
  year={2009},
  publisher={SAGE Publications Limited}
}

@article{marques1999grasp,
  title={GRASP: A search algorithm for propositional satisfiability},
  author={Marques-Silva, Joao P and Sakallah, Karem A},
  journal={IEEE Transactions on computers},
  volume={48},
  number={5},
  pages={506--521},
  year={1999},
  publisher={IEEE}
}

@inproceedings{moskewicz2001chaff,
  title={Chaff: Engineering an efficient SAT solver},
  author={Moskewicz, Matthew W and Madigan, Conor F and Zhao, Ying and Zhang, Lintao and Malik, Sharad},
  booktitle={Proceedings of the 38th annual Design Automation Conference},
  pages={530--535},
  year={2001}
}

@article{queue2019cadical,
  title={CaDiCaL at the SAT Race 2019},
  author={Queue, Separate DECISION},
  journal={SAT RACE},
  volume={2019},
  pages={8},
  year={2019}
}

@article{Hale1980,
	author    = {Hale, William K.},
	title     = {Frequency assignment: Theory and applications},
	journal   = {Proceedings of the IEEE},
	volume    = {68},
	number    = {12},
	pages     = {1497--1514},
	year      = {1980},
	doi       = {10.1109/PROC.1980.11899}
}

@article{GriggsYeh1992,
	author    = {Griggs, Jerrold R. and Yeh, Roger K.},
	title     = {Labeling graphs with a condition at distance~2},
	journal   = {SIAM Journal on Discrete Mathematics},
	volume    = {5},
	number    = {4},
	pages     = {586--595},
	year      = {1992},
	doi       = {10.1137/0405048}
}

@article{Chartrand2001,
	author    = {Chartrand, Gary and Erwin, David and Harary, Frank and Zhang, Ping},
	title     = {Radio labelings of graphs},
	journal   = {Bulletin of the Institute of Combinatorics and its Applications},
	volume    = {33},
	pages     = {77--85},
	year      = {2001}
}

@article{LiuZhu2005,
	author    = {Liu, Daphne Der-Fen and Zhu, Xuding},
	title     = {Multilevel distance labelings for paths and cycles},
	journal   = {SIAM Journal on Discrete Mathematics},
	volume    = {19},
	number    = {3},
	pages     = {610--621},
	year      = {2005},
	doi       = {10.1137/S0895480102417768}
}

@article{LiuXie2004,
	author    = {Liu, Daphne Der-Fen and Xie, Minhui},
	title     = {Radio number for square of cycles},
	journal   = {Congressus Numerantium},
	volume    = {169},
	pages     = {105--125},
	year      = {2004}
}

@article{LiuXie2009,
	author    = {Liu, Daphne Der-Fen and Xie, Minhui},
	title     = {Radio number for square of paths},
	journal   = {Ars Combinatoria},
	volume    = {90},
	pages     = {307--319},
	year      = {2009}
}

@article{Liu2008,
	author    = {Liu, Daphne Der-Fen},
	title     = {Radio number for trees},
	journal   = {Discrete Mathematics},
	volume    = {308},
	number    = {7},
	pages     = {1153--1164},
	year      = {2008},
	doi       = {10.1016/j.disc.2007.03.066}
}

@article{LiMakZhou2010,
	author    = {Li, Xiangwen and Mak, Vicky and Zhou, Sanming},
	title     = {Optimal radio labellings of complete $m$-ary trees},
	journal   = {Discrete Applied Mathematics},
	volume    = {158},
	number    = {5},
	pages     = {507--515},
	year      = {2010},
	doi       = {10.1016/j.dam.2009.11.014}
}

@article{Kchikech2007,
	author    = {Kchikech, Mustapha and Khennoufa, Riadh and Togni, Olivier},
	title     = {Linear and cyclic radio $k$-labelings of trees},
	journal   = {Discussiones Mathematicae Graph Theory},
	volume    = {27},
	number    = {1},
	pages     = {105--123},
	year      = {2007},
	doi       = {10.7151/dmgt.1348}
}

@article{ReddyIyer2011,
	author    = {VenkataSubbaReddy, P. and ViswanathanIyer, K.},
	title     = {Upper bounds on the radio number of some trees},
	journal   = {International Journal of Pure and Applied Mathematics},
	volume    = {71},
	number    = {2},
	pages     = {207--215},
	year      = {2011}
}

@article{Saha2022,
	author    = {Saha, Laxman and Basunia, Alamgir Rahaman and Das, Sandip and others},
	title     = {Radio $k$-chromatic number of full $m$-ary trees},
	journal   = {Theory of Computing Systems},
	volume    = {66},
	pages     = {114--142},
	year      = {2022},
	doi       = {10.1007/s00224-021-10056-7}
}

@article{KhennoufahTogni2011,
	author    = {Khennoufa, Riadh and Togni, Olivier},
	title     = {The radio antipodal and radio numbers of the hypercube},
	journal   = {Ars Combinatoria},
	volume    = {102},
	pages     = {447--461},
	year      = {2011}
}

@article{Ortiz2011,
	author    = {Ortiz, J. P. and Martinz, P. and Tomova, M. and Wyels, C.},
	title     = {Radio number of some generalized prism graphs},
	journal   = {Discussiones Mathematicae Graph Theory},
	volume    = {31},
	number    = {1},
	pages     = {45--62},
	year      = {2011},
	doi       = {10.7151/dmgt.1530}
}

@article{KhennoufahTogni2005,
	author    = {Khennoufa, Riadh and Togni, Olivier},
	title     = {A note on radio antipodal colorings of paths},
	journal   = {Mathematica Bohemica},
	volume    = {130},
	number    = {3},
	pages     = {277--282},
	year      = {2005}
}

@article{KolaPanigrahi2009a,
	author    = {Kola, Srinivasa Rao and Panigrahi, Pratima},
	title     = {Nearly antipodal chromatic number $\mathrm{ac}'(P_n)$
	of a path~$P_n$},
	journal   = {Mathematica Bohemica},
	volume    = {134},
	number    = {1},
	pages     = {77--86},
	year      = {2009}
}

@article{JuanLiu2012,
	author    = {Juan, Jou-Shin T. and Liu, Daphne Der-Fen},
	title     = {Antipodal labelings for cycles},
	journal   = {Ars Combinatoria},
	volume    = {103},
	pages     = {81--96},
	year      = {2012}
}

@article{KolaPanigrahi2015,
	author    = {Kola, Srinivasa Rao and Panigrahi, Pratima},
	title     = {A lower bound for radio $k$-chromatic number of an arbitrary graph},
	journal   = {Contributions to Discrete Mathematics},
	volume    = {10},
	number    = {2},
	pages     = {45--56},
	year      = {2015}
}

@article{Das2017,
	author    = {Das, Sandip and Ghosh, Sasthi C. and Nandi, Soumen and Sen, Sagnik},
	title     = {A lower bound technique for radio $k$-coloring},
	journal   = {Discrete Mathematics},
	volume    = {340},
	number    = {5},
	pages     = {855--861},
	year      = {2017},
	doi       = {10.1016/j.disc.2016.12.021}
}

@article{SahaPanigrahi2015,
	author    = {Saha, Laxman and Panigrahi, Pratima},
	title     = {A lower bound for radio $k$-chromatic number},
	journal   = {Discrete Applied Mathematics},
	volume    = {192},
	pages     = {87--100},
	year      = {2015},
	doi       = {10.1016/j.dam.2014.05.037}
}

@article{Chakraborty2024,
	author    = {Chakraborty, Dipayan and Nandi, Soumen and Sen, Sagnik
	and Supraja, D. K.},
	title     = {A linear algorithm for radio $k$-coloring of powers of
	paths having small diameters},
	journal   = {Journal of Computer and System Sciences},
	volume    = {146},
	pages     = {103577},
	year      = {2024},
	doi       = {10.1016/j.jcss.2024.103577}
}

@article{ELrokh2022,
	author    = {ELrokh, Ashraf and Badr, Elsayed M. and Al-Shamiri,
	Mohammed M. Ali and Ramadhan, Shimaa},
	title     = {Upper bounds of radio number for triangular snake and
	double triangular snake graphs},
	journal   = {Journal of Mathematics},
	volume    = {2022},
	pages     = {3635499},
	year      = {2022},
	doi       = {10.1155/2022/3635499}
}

@article{Alkasasbeh2023,
	author    = {Alkasasbeh, Ahmad H. and Badr, Elsayed M. and Attiya,
	Heba and others},
	title     = {Radio number for friendship communication networks},
	journal   = {Mathematics},
	volume    = {11},
	number    = {1},
	pages     = {1--14},
	year      = {2023},
	doi       = {10.3390/math11010001}
}

@article{KimSong2015,
	author    = {Kim, Byeong Moon and Song, Byung Chul and Hwang, Woonjae},
	title     = {Radio number for the product of a path and a complete graph},
	journal   = {Journal of Combinatorial Optimization},
	volume    = {30},
	number    = {1},
	pages     = {139--149},
	year      = {2015},
	doi       = {10.1007/s10878-013-9639-3}
}

@article{BantvaLiu2021,
	author    = {Bantva, Devsi and Liu, Daphne Der-Fen},
	title     = {Optimal radio labellings of block graphs and line graphs
	of trees},
	journal   = {Theoretical Computer Science},
	volume    = {891},
	pages     = {90--104},
	year      = {2021},
	doi       = {10.1016/j.tcs.2021.09.022}
}

@article{VasoyaBantva2024,
    author = {Vasoya, Payal},
    year = {2025},
    month = {01},
    pages = {47-64},
    title = {Radio number of the cartesian product of a tree and a complete graph},
    volume = {235},
    journal = {Congressus Numerantium},
    doi = {10.61091/cn235-05}
}

@article{VasoyaBantva2023,
	author    = {Vasoya, Payal and Bantva, Devsi},
	title     = {Optimal radio labelings of the {Cartesian} product of the
	generalized {Petersen} graph and tree},
	journal   = {Discrete Mathematics, Algorithms and Applications},
	volume    = {16},
	pages     = {1--15},
	year      = {2024},
	doi       = {10.1142/S1793830923500258}
}

@article{SarkarAdhikari2015,
	author    = {Sarkar, Ushnish and Adhikari, Avishek},
	title     = {On characterizing radio $k$-coloring problem by path
	covering problem},
	journal   = {Discrete Mathematics},
	volume    = {338},
	number    = {4},
	pages     = {615--620},
	year      = {2015},
	doi       = {10.1016/j.disc.2014.10.013}
}

@InProceedings{bantva2019,
author={Bantva, Devsi},
editor={Pal, Sudebkumar Prasant
and Vijayakumar, Ambat},
title={A Lower Bound for the Radio Number of Graphs},
booktitle="Algorithms and Discrete Applied Mathematics (CALDAM 2019)",
series = {Lecture Notes in Computer Science},
year="2019",
publisher="Springer International Publishing",
address="Cham",
pages="161--173",
doi = {10.1007/978-3-030-11509-8_14}
}

@inproceedings{ignatiev2018pysat,
  title={PySAT: A Python toolkit for prototyping with SAT oracles},
  author={Ignatiev, Alexey and Morgado, Antonio and Marques-Silva, Joao},
  booktitle={International Conference on Theory and Applications of Satisfiability Testing},
  pages={428--437},
  year={2018},
  organization={Springer}
}
	
\end{document}